\documentclass[leqno,a4paper,12pt]{article}

\usepackage{amsthm,amsmath,amssymb,upref} 
\usepackage{epsfig}
\newcommand{\scale}{1.2} 

\renewcommand{\phi}{\varphi}
\renewcommand{\theta}{\vartheta}
\renewcommand{\epsilon}{\varepsilon}
\DeclareMathSymbol{\Gamma}{\mathalpha}{letters}{"00}
\DeclareMathSymbol{\Delta}{\mathalpha}{letters}{"01}
\DeclareMathSymbol{\Theta}{\mathalpha}{letters}{"02}
\DeclareMathSymbol{\Lambda}{\mathalpha}{letters}{"03}
\DeclareMathSymbol{\Xi}{\mathalpha}{letters}{"04}
\DeclareMathSymbol{\Pi}{\mathalpha}{letters}{"05}
\DeclareMathSymbol{\Sigma}{\mathalpha}{letters}{"06}
\DeclareMathSymbol{\Upsilon}{\mathalpha}{letters}{"07}
\DeclareMathSymbol{\Phi}{\mathalpha}{letters}{"08}
\DeclareMathSymbol{\Psi}{\mathalpha}{letters}{"09}
\DeclareMathSymbol{\Omega}{\mathalpha}{letters}{"0A}

\theoremstyle{plain}
\newtheorem{theorem}{Theorem}[section]
\newtheorem{lemma}[theorem]{Lemma}
\newtheorem{corollary}[theorem]{Corollary}
\newtheorem{proposition}[theorem]{Proposition}

\theoremstyle{definition}

\newtheorem{example}[theorem]{Example}
\newtheorem{remark}[theorem]{Remark}

\numberwithin{equation}{section}

\newcommand{\theoremlabel}[1]{\label{thm:#1}}
\newcommand{\theoremref}[1]{\textup{Theorem~\ref{thm:#1}}}
\newcommand{\lemmalabel}[1]{\label{lem:#1}}
\newcommand{\lemmaref}[1]{\textup{Lemma~\ref{lem:#1}}}
\newcommand{\corollarylabel}[1]{\label{cor:#1}}
\newcommand{\corollaryref}[1]{\textup{Corollary~\ref{cor:#1}}}
\newcommand{\sectionlabel}[1]{\label{sec:#1}}
\newcommand{\sectionref}[1]{\textup{Section~\ref{sec:#1}}}
\newcommand{\appendixlabel}[1]{\label{etc:#1}}
\newcommand{\appendixref}[1]{\textup{Appendix~\ref{etc:#1}}}
\newcommand{\examplelabel}[1]{\label{exa:#1}}
\newcommand{\exampleref}[1]{\textup{Example~\ref{exa:#1}}}

\newcommand{\propositionlabel}[1]{\label{prp:#1}}
\newcommand{\propositionref}[1]{\textup{Proposition~\ref{prp:#1}}}
\newcommand{\remarklabel}[1]{\label{rem:#1}}
\newcommand{\remarkref}[1]{\textup{Remark~\ref{rem:#1}}}

\newcommand{\R}{{R}}
\newcommand{\onehalf}{\tfrac12}
\newcommand{\trans}{\mathsf{T}}
\newcommand{\Gt}{\tilde{G}}
\newcommand{\Nt}{\tilde{N}}
\newcommand{\Wt}{\tilde{W}}
\newcommand{\dG}{\det\Gt}
\newcommand{\grad}{\nabla}
\DeclareMathOperator{\cof}{cof}
\DeclareMathOperator{\diag}{diag}
\DeclareMathOperator{\tr}{tr}
\newcommand{\sigmacheck}[2]{\sigma_{#1}(\check{u}_{#2})}
\newcommand{\Jac}{J}
\newcommand{\abs}[1]{\lvert#1\rvert}

\title{St\"ackel separability for \\
Newton systems of cofactor type\thanks{September 2003.}}

\author{Stefan Rauch-Wojciechowski\thanks{Matematiska institutionen, Link\"opings universitet, \hbox{SE-581~83} Link\"oping, Sweden (\texttt{strau@mai.liu.se}, \texttt{clwak@mai.liu.se}).} 
\and Claes Waksj\"o\footnotemark[2]}

\date{}

\begin{document}
\maketitle

\begin{abstract} 
A conservative Newton system $\ddot{q}=-\grad V(q)$ in $\R^n$ is called separable when the Hamilton--Jacobi equation for the natural Hamiltonian $H=\onehalf p^2+V(q)$ can be solved through separation of variables in some curvilinear coordinates. If these coordinates are orhogonal, the Newton system admits $n$ first integrals, which all have separable St\"ackel form with quadratic dependence on $p$.

We study here separability of the more general class of Newton systems $\ddot{q}=-(\cof G)^{-1}\grad W(q)$ that admit $n$ quadratic first integrals. We prove that a \emph{related} system with the same integrals can be transformed through a \emph{non-canonical} transformation into a St\"ackel separable Hamiltonian system and solved by qudratures, providing a solution to the original system.

The separation coordinates, which are defined as characteristic roots of a linear pencil $G-\mu\Gt$ of elliptic coordinates matrices, generalize the well known elliptic and parabolic coordinates. Examples of such new coordinates in two and three dimensions are given.

These results extend, in a new direction, the classical separability theory for natural Hamiltonians developed in the works of Jacobi, Liouville, St\"ackel, Levi-Civita, Eisenhart, Benenti, Kalnins and Miller.

\textbf{Key words.} Separability, Hamilton--Jacobi equation, integrability, cofactor pair systems.
\end{abstract}

\clearpage\tableofcontents\clearpage


\section{Introduction}

The method of separation of variables for natural Hamiltonian systems is well understood~\cite{jac,sb-cc-gr}. In terms of (orthogonal) separation coordinates, the natural Hamiltonian $H(q,p)=\onehalf p^{2}+V(q)$ becomes
\begin{equation}
\label{natham}
\tilde{H}(x,y)=H(q(x),p(x,y))=
\onehalf\sum_{i=1}^{n}H_i^{-2}(x)\,y_i^2+V(q(x)),
\end{equation}
and separation means that its Hamilton--Jacobi equation, obtained by substituting $y_i=\partial S(x;\alpha)/\partial x_i$, admits an additively separated solution $S(x;\alpha)=\sum_{i=1}^nS_i(x_i;\alpha)$. After this substitution the
Hamilton--Jacobi equation splits into a system of first order ODEs for the functions $S_i(x_i;\alpha)$ and the solution can be expressed by qudratures.

The well known St\"ackel theorem~\cite{stackhab,stackann} gives a necessary and sufficient conditions for $H_1^{-2}$, \dots, $H_n^{-2}$ and $V(q(x))$ to admit a separated solution of the Hamilton--Jacobi equation. This St\"ackel condition is purely algebraic, and we say that the Hamiltonian~\eqref{natham} has separable St\"ackel form if these conditions are fulfilled. The change of coordinates $q(x)$ to separation coordinates $x$ is a priori not known and the problem of finding these variables for any given potential $V(q)$ has been solved only recently~\cite{cw-srw}.

\medskip

For general dynamical systems $\dot{x}=f(x)$, $x=(x_1,\dots,x_n)^\trans$, or for non-canonical Hamiltonian systems $\dot{z}=\Pi(z)\nabla H(z)$, where $\Pi(z)$ is a Poisson matrix, the notion of Hamilton--Jacobi separability is not well understood. The reason is that for non-canonical Hamiltonian systems there is no naturally associated Hamilton--Jacobi equation that can be solved by an additive ansatz $S=\sum_{i=1}^nS_i$.

In this paper we study systems of autonomous second order ordinary
differential equations of the form $\ddot{q}=M(q)$, $q=(q_1,\dots,q_n)^\trans$,  where acceleration $\ddot{q}$ equals a force $M(q)$, which does not depend on velocities~$\dot{q}$. Such systems have been given the name Newton systems. We consider Newton systems of a special type having a force $M(q)$ that is quasi-potential in two different ways $M(q)=-(\cof G)^{-1}\grad W(q)=-(\cof\Gt)^{-1}\grad \Wt(q)$, where $G$, $\Gt$ are elliptic coordinates matrices, $\cof G=(\det G)G^{-1}$ and $W(q)$, $\Wt(q)$ are quasi-potentials. Such cofactor pair systems naturally generalize separable potential Newton systems $\ddot{q}=-\grad V(q)$ in $\R^n$ since the function $\Wt(q)$ becomes an ordinary potential for $\Gt=I$.

Any cofactor pair system can be written as a dynamical system $\dot{q}=p$, $\dot{p}=M(q)$, and is known to possess, in the extended phase space, a non-canonical Hamiltonian formulation $\dot{z}=\Pi(z)\grad H(z)$, $z=(q_1,\dots,q_n,p_1,\dots,p_n,d)^\trans$. These Newton systems have been proved to be integrable in a somewhat non-standard way through embedding into a Liouville integrable system \cite{hl,srw-km-hl}.

In the first part of this paper we show that any cofactor pair system $\ddot{q}=-(\cof G)^{-1}\grad W(q)=-(\cof\Gt)^{-1}\grad \Wt(q)$ has a related system that can be transformed through a non-canonical transformation into a canonical Hamiltonian system
\begin{displaymath}
\dot{x}=\frac{\partial H(x,y)}{\partial y}, \quad 
\dot{y}=-\frac{\partial H(x,y)}{\partial x},
\end{displaymath}
which has separable St\"{a}ckel form (\theoremref{cofseparability}). A solution of the related system then provides a solution of the cofactor pair system by rescaling the time variable. This result indicates a simple and natural way of prescribing the Hamilton--Jacobi separability property to dynamical systems that are not canonical Hamiltonian systems. Such systems can be considered to be separable whenever there exists a transformation (usually not a canonical one) into a canonical Hamiltonian system that admits the classical Hamilton--Jacobi separability (see \remarkref{srwnonham} and \remarkref{srwnoncan}). This observation allows for extending the concept of separability and the concept of solving through separation of variables to large classes of dynamical systems that otherwise could not be considered separable in the Hamilton--Jacobi sense.

Separability for generic cofactor pair systems has already been studied \cite{km-mb}, and separation coordinates in the extended phase space of variables $(q,p,d)$ have been derived using the theory of bi-Hamiltonian systems developed by the Milano school~\cite{fm-gf-mp}. Separability for the more degenerate class of driven cofactor systems has also been studied \cite{hl-srw}. In this paper we approach the problem of separability without using the concept of extended phase space and find a direct transformation into separation coordinates in which the related system attains separable St\"ackel form. The advantage of this approach is that it explains, on the level of differential equations, the mechanism of separability and gives simple tools for solving these systems without resorting to the geometrical language of the Milano school.

The second part of this paper studies cofactor-elliptic coordinates defined as
characteristic roots of a linear pencil $G-\mu\Gt$ of two elliptic coordinates matrices $G$, $\Gt$. They separate cofactor pair systems and appear to be a natural, however more complicated, generalisation of the elliptic coordinates. 

Elliptic coordinates $x_1$, \dots, $x_n$ play a central role in the classical separability theory for natural Hamiltonians~\eqref{natham}. They are defined as zeroes of
\begin{displaymath}
1+\sum_{i=1}^n\frac{q_i^2}{z-\lambda_i}=
\prod_{j=1}^n(z-x_j)\biggm/\prod_{k=1}^n(z-\lambda_k)
\end{displaymath}
where $\lambda_1$, \dots, $\lambda_n$ are parameters. The elliptic coordinates are mother of all orthogonal separable coordinates on $\R^n$ for the natural Hamilton-Jacobi equations. This means that all other separation coordinates can be recovered from the elliptic coordinates through certain proper or improper degenerations of the parameters $\lambda_i$. For cofactor-elliptic coordinates we find a formula [see \eqref{cpsell}], which closely reminds of the above formula but encompasses considerably more coordinate systems. These coordinates are usually not orthogonal in the ordinary Euclidean sense, but instead orthogonal with respect to the scalar product $(v,w)\mapsto v^\trans\Gt w$. For instance, in two dimensions, cofactor elliptic coordinates $x_1$, $x_2$ are defined as solutions of
\begin{displaymath}
\frac{(q_1+\epsilon x)^{2}}{\lambda_1-(1-\epsilon)x+\epsilon^2x^2}
+\frac{q_2^2}{\lambda_2-x}=1.
\end{displaymath}
For $\epsilon=0$ this equation defines two-dimensional elliptic coordinates. Thus $\lambda_1$, $\lambda_2$ play a similar role as before, while for other values of the parameter~$\epsilon$, the curves of constant value of $x$ cover the plane in a complicated way. Their pattern strongly depends on the relative values of $\lambda_1$, $\lambda_2$, and usually not all points of the plane are parametrised through cofactor elliptic coordinates $x_1$, $x_2$. In such domains separation takes place for complex values of $x_1$, $x_2$. Several examples of these coordinates are given and illustrated by pictures.

In this paper we discuss and exemplify only the generic type cofactor-elliptic and cofactor-parabolic coordinates that play a fundamental role for cofactor pair systems. This study is the first step into a new fascinating world of nonorthogonal separation coordinates defined by families of non-confocal quadrics.

\subsection{Properties of cofactor pair systems}

Before discussing separability of cofactor pair systems, we need to define them and recall some facts from~\cite{hl}.

An elliptic coordinates matrix is a matrix-valued function of $q\in\R^n$ that can be written in the form 
\begin{displaymath}
 G(q) = \alpha qq^\trans+q\beta^\trans+\beta q^\trans+\gamma
\end{displaymath}
[we refer to elements $q=(q_1,\dots,q_n)^\trans$ of $\R^n$ as vectors and view them as $n\times1$ matrices], where $\alpha\in\R$ is a constant scalar, $\beta\in\R^n$ is a constant vector and $\gamma=\gamma^\trans$ is a real constant symmetric $n\times n$ matrix. By construction, $G$ is also symmetric.

A simple but very useful fact is that an arbitrary linear combination of elliptic coordinates matrices is an elliptic coordinates matrix too. This implies that all statements to be made about elliptic coordinates matrices also hold for pencils $G-\mu\Gt$ of elliptic coordinates matrices.

For a quadratic matrix $A$, we write $\cof A$ to denote the corresponding cofactor matrix, which is defined by $(\cof A)A=A(\cof A)=(\det A)I$.

To a given elliptic coordinates matrix $G$, there is an associated vector 
\begin{displaymath}
N=\alpha q+\beta=\onehalf\grad\tr G
\end{displaymath}
[$\grad$ denotes the gradient operator $\partial_q=(\partial/\partial q_1,\dots,\partial/\partial q_n)^\trans$ expressed in Cartesian coordinates]. If $G$ is non-singular, it is possible to represent this vector as 
\begin{equation}
\label{assvecrepr}
N=\onehalf G\grad\log\det G
\end{equation}
[we use logarithmic derivatives like $\grad\log F$ formally to denote $(1/F)\grad F$ regardless of the sign of $F$]. The associated vector satisfies \begin{equation}
\label{assvecid}
\grad\bigl(N^\trans(\cof G)N\bigr)=2\alpha(\cof G)N.
\end{equation}

A Newton system $\ddot{q}=M(q)$ in $\R^n$ is called a cofactor pair system if the force can be generated in two different ways as
\begin{displaymath}
 M(q) = -(\cof G)^{-1}\grad W = -(\cof\Gt)^{-1}\grad\Wt,
\end{displaymath}
where 
\begin{align*}
 G(q) &= \alpha qq^\trans+q\beta^\trans+\beta q^\trans+\gamma,&
 \det G &\ne 0,\\
 \Gt(q) &= \tilde{\alpha}qq^\trans+q\tilde{\beta}^\trans+\tilde{\beta} 
 q^\trans+\tilde{\gamma},&
 \det\Gt &\ne 0,
\end{align*}
are two linearly independent nonsingular elliptic coordinates matrices. Here $W$ and $\Wt$ are two functions on $\R^n$ called quasi-potentials. Clearly, these quasi-potentials have to satisfy 
\begin{equation}
\label{prefe}
(\cof G)^{-1}\grad W = (\cof\Gt)^{-1}\grad\Wt.
\end{equation}
The Frobenius compatibility conditions for this system of equations, rewritten in terms of $K=W/\det G$ and $\tilde{K}=\Wt/\det\Gt$, are referred to as the fundamental equations.

By defining 
\begin{displaymath}
 E=\onehalf\dot{q}^\trans(\cof G)\dot{q}+W\quad\text{and}\quad
 \tilde{E}=\onehalf\dot{q}^\trans(\cof\Gt)\dot{q}+\Wt,
\end{displaymath}
two quadratic first integrals of ``energy type'' are constructed. In fact, the so-called ``$2\Rightarrow n$ theorem'' states that any cofactor pair system admits $n$ quadratic first integrals
\begin{displaymath}
 E^{(k)}=\onehalf\dot{q}^\trans A^{(k)}\dot{q}+W^{(k)},\quad
 k=0,\dots,n-1,
\end{displaymath}
where the matrices $A^{(k)}$ are given by the generating function 
\begin{displaymath}
\cof(G+\mu\Gt)=\sum_{k=0}^{n-1}\mu^k A^{(k)}.
\end{displaymath}
The quasi-potentials $W^{(k)}$ are given as solutions to the differential equations 
\begin{displaymath}
\grad W^{(k)}=-A^{(k)}M,
\end{displaymath}
which are integrable if the fundamental equations are satisfied by $K=W/\det G$ and $\tilde{K}=\Wt/\det\Gt$. In particular, $E^{(0)}=E$ and $E^{(n-1)}=\tilde{E}$.

In this paper, we focus on cofactor pair systems having the property that the polynomial $\det(G-\mu\Gt)$ has $n$ functionally independent roots $\mu=u_k(q)$. We call such systems \emph{generic.} In the generic situation, the functions $u_k(q)$ are necessarily non-constant and give rise to the non-singular Jacobian
\begin{displaymath}
 \Jac=(\grad u_1,\dots,\grad u_n),\quad\det\Jac\ne0.
\end{displaymath}
It is thus clear that we can define new coordinates as being the roots $u_k=u_k(q)$. For this change of variables we can write the chain rule compactly as $\grad=\Jac\partial_u$. The gradients $\grad u_k$ are $\Gt$-orthogonal, $(\grad u_j)^\trans\Gt\grad u_k=0$ for $j\ne k$, and give rise to generalized metric coefficients $\Delta_k=(\grad u_k)^\trans\Gt\grad u_k$. Thus we have the generalized orthogonality relation
\begin{equation}
\label{genort}
\Jac^\trans\Gt\Jac=\Delta=\diag(\Delta_1,\dots,\Delta_n).
\end{equation}
 
Writing 
\begin{displaymath}
X=\Gt^{-1}G,
\end{displaymath}
we see from $G-\mu\Gt=\Gt(X-\mu I)$ that $u_k$ are eigenvalues of $X$. The corresponding eigenvectors are the gradients $\grad u_k$, so that $X\grad u_k=u_k\grad u_k$ for all $k$, which we can write as
\begin{equation}
\label{eigrel}
X\Jac=\Jac\mathcal{U},\quad\mathcal{U}=\diag(u_1,\dots,u_n).
\end{equation}
This implies that $X$ satisfies 

\begin{equation}
\label{xid}
X\grad\log\det X=\grad\tr X.
\end{equation}

\section{Main separability theorem}
 
In~\cite{hl} it was shown that to any cofactor pair system $\ddot{q}=M(q)$ there is a related bi-Hamiltonian system
\begin{displaymath}
        \frac{d}{d\tau}
        \begin{pmatrix}
                q \\ p \\ d
        \end{pmatrix}
        =
        \begin{pmatrix}
                0 & \onehalf G & p \\
                -\onehalf G^\trans & \onehalf(Np^\trans-pN^\trans) & M+dN \\
                -p^\trans & -(M+dN)^\trans & 0
        \end{pmatrix}
        \begin{pmatrix}
                \partial_q \\ \partial_p \\ \partial_d
        \end{pmatrix}
        (d\dG)
\end{displaymath}
in the extended $(2n+1)$-dimensional phase space obtained by taking $p=\dot{q}$ and introducing an extra variable $d$. On the hyperplane $d=0$, the bi-Hamiltonian system reduces to 
\begin{equation}
\label{qpdrel}
        \frac{d}{d\tau}
        \begin{pmatrix}
                q \\ p \\ d
        \end{pmatrix}
        =(\dG)
        \begin{pmatrix}
                p \\ M \\ 0
        \end{pmatrix}.
\end{equation}
This system has the same trajectories as the system
\begin{equation}
\label{qpd}
        \frac{d}{dt}
        \begin{pmatrix}
                q \\ p \\ d
        \end{pmatrix}
        =
        \begin{pmatrix}
                p \\ M \\ 0
        \end{pmatrix},
\end{equation}
which clearly is equivalent to $\ddot{q}=M(q)$ when $d=0$. Since the trajectories are the same, the solutions to \eqref{qpdrel} and \eqref{qpd} differ only by a scaling of the independent variable. So if the solution of \eqref{qpdrel} is known, the solution of \eqref{qpd} can be obtained by substituting the correct expression $\tau(t)$ for~$\tau$. In this sense, every cofactor pair system is equivalent to a Hamiltonian system. 

Using these ideas, it has been shown that any cofactor pair system is Liouville integrable (in a somewhat non-standard sense), provided that the first integrals in the $2\Rightarrow n$ theorem are functionally independent and in involution.

For cofactor pair systems, explicit integration through the Hamilton--Jacobi method has also been shown~\cite{km-mb}. Guided by the theory of Darboux--Nijenhuis coordinates for bi-Hamiltonian systems, the separation coordinates $(u,s,c)$ were introduced, where $u_k(q)$ are defined as roots of the polynomial $\det(G+\mu\Gt)$, while
\begin{equation}
\label{sdef}
        s_k(q,p)=\frac
        {\Omega^\trans\cof\bigl(G+u_k(q)\Gt\bigr)p}
        {\Omega^\trans\cof\bigl(G+u_k(q)\Gt\bigr)\Omega},\quad
        \Omega=G\Gt^{-1}\Nt-N,
\end{equation}
and $c=(\dG)d$. In these coordinates, the related system \eqref{qpdrel} takes the form
\begin{displaymath}
        \frac{d}{d\tau}
        \begin{pmatrix}
                u \\ s \\ c
        \end{pmatrix}
        =
        \begin{pmatrix}
                0 & I & 0 \\
                -I & 0 & 0 \\
                0 & 0 & 0
        \end{pmatrix}
        \begin{pmatrix}
                \partial_u \\ \partial_s \\ \partial_c
        \end{pmatrix}
        h(u,s,c),
\end{displaymath}
and the Hamiltonian $h(u,s,c)$ attains separable St\"ackel form. When $c=0$ one then obtains a $2n$-dimensional canonical Hamiltonian system in $(u,s)$ variables, so the classical theory is applicable. The connection between the $(2n+1)$-dimensional system and the cofactor pair system $\ddot{q}=M(q)$ is however intricate and it is not easy to see the separation mechanism for cofactor pair systems. 

We will now give an alternative formulation of this result, in the spirit of the classical St\"ackel approach. The techniques used in the proof are further extensions of the results in~\cite{hl-srw}.

\begin{theorem}[St\"ackel separability for generic cofactor pair systems] 
\theoremlabel{cofseparability}
Let $d^2q/dt^2=M(q)=-(\cof G)^{-1}\grad W=-(\cof\Gt)^{-1}\grad\Wt$ be a cofactor pair system in $\R^n$ written as a dynamical system
 \begin{equation}
        \label{originalsystem}
        \frac{d}{dt}
        \begin{pmatrix}
                q \\ p
   \end{pmatrix}
        = 
        \begin{pmatrix}
                p \\ M
 \end{pmatrix}
 \end{equation}
with the related system
 \begin{equation}
        \label{relatedsystem}
        \frac{d}{d\tau}
        \begin{pmatrix}
                q \\ p
   \end{pmatrix}
        = (\dG)
        \begin{pmatrix}
                p \\ M
   \end{pmatrix}
 \end{equation}
obtained by rescaling the vector field \eqref{originalsystem}. 
 
Suppose that the polynomial $\det(G-\mu\Gt)$ has $n$ functionally independent roots $\mu=u_k(q)$, so that they define new configuration coordinates $u_k=u_k(q)$. Define new momenta coordinates as
 \begin{equation}
 \label{momenta}
  s_k=s_k(q,p)=
  \frac{1}{\Delta_k}\sum_{i=1}^n\frac{\partial u_k}{\partial q_i}p_i
 \end{equation}
where $\Delta_k=(\grad u_k)^\trans\Gt\grad u_k$ are generalized metric coefficients. In these new coordinates we have:
 
 \begin{enumerate}
        \item The related system \eqref{relatedsystem} has a canonical 
        Hamiltonian formulation
        \begin{equation}
    \label{canhamsys}
         \frac{d}{d\tau}
         \begin{pmatrix}
                u \\ s
         \end{pmatrix}
         =
         \begin{pmatrix}
                0 & I \\ -I & 0
         \end{pmatrix}
         \begin{pmatrix}
                \partial_u \\ \partial_s
         \end{pmatrix}
         H
        \end{equation}
        with Hamiltonian
        \begin{equation}
        \label{hamiltonian}
         H=\onehalf(\dG)\sum_{i=1}^n\Delta_is_i^2+\Wt(q(u)).
        \end{equation}
        
        \item The Hamiltonian \eqref{hamiltonian} has separable St\"ackel form
        \begin{displaymath}
         H=\sum_{i=1}^n
         \frac{\sigmacheck{0}i}{U'(u_i)}
         \bigl(\onehalf f_i(u_i)\,s_i^2+g_i(u_i)\bigr),
        \end{displaymath}
    where $f_i$ and $g_i$ are some functions of one variable. 
        The St\"ackel matrix is the Vandermonde matrix \eqref{vandermonde}
        and consequently, the coefficients $\sigmacheck{0}i/U'(u_i)$ are 
        elements of its inverse (see \appendixref{sigma}). 
        
        \item Systems \eqref{originalsystem} and 
        \eqref{relatedsystem} have a common set of $n$
        quadratic first integrals 
        \begin{displaymath}
         E^{(k)}=\sum_{i=1}^n
         \frac{\sigmacheck{n-k-1}i}{U'(u_i)}
         \bigl(\onehalf f_i(u_i)\,s_i^2+g_i(u_i)\bigr),
         \quad k=0,\dots,n-1,
        \end{displaymath}
        where $f_i$ and $g_i$ are the same functions as above.
        These first integrals are 
        functionally independent and in involution.
        The Hamiltonian \eqref{hamiltonian} is given by $H=E^{(n-1)}$.

        \item The Hamilton--Jacobi equations
        $E^{(k)}(u,\partial_uS)=\alpha_k$, $k=0$, \dots, $n-1$, 
        admit a common separated solution $S(u;\alpha)=\sum S_k(u_k;\alpha)$. 
        In particular, $S$ is a solution to the last equation
        \begin{displaymath}
         \onehalf(\dG)\sum_{i=1}^n\Delta_i
         \left(\frac{\partial S}{\partial u_i}\right)^2
         +\Wt(q(u))=\alpha_{n-1}
        \end{displaymath}       
        associated to the Hamiltonian \eqref{hamiltonian}. The corresponding
        Hamiltonian system \eqref{canhamsys} and the cofactor pair system
        \eqref{originalsystem} can therefore be solved by quadratures.
 \end{enumerate}
\end{theorem}

We shall clarify the content of this theorem and the connection with classical separability of natural Hamiltonian systems in the subsequent remarks.

\begin{remark}[Change of coordinates]
It is convenient to rewrite formula~\eqref{momenta} for the change of momenta coordinates  in matrix notation. With the previously introduced generalized metric matrix $\Delta$ and the Jacobian $\Jac$, we have 
\begin{displaymath}
s=\Delta^{-1}\Jac^\trans p.
\end{displaymath}
It is easy to see that if $\Jac$ is invertible, then so is the full phase space transformation $(q,p)\to(u,s)$.
\end{remark}

\begin{remark}[Definition of momenta]
One can show that definition \eqref{momenta} agrees with \eqref{sdef}. The form \eqref{momenta} is however more suitable for our proof.
\end{remark}
        
\begin{remark}[Comparison with the classical case]
Any cofactor system $\ddot{q}=-(\cof\Gt)^{-1}\grad\Wt$ admits a first integral of energy type, $\tilde{E}=\onehalf\dot{q}^\trans(\cof\Gt)\dot{q}+\Wt$, by construction. Stated in a non-symmetric way, \theoremref{cofseparability} says that a cofactor system $\ddot{q}=-(\cof\Gt)^{-1}\grad\Wt$ is separable if there is an extra first integral, also of energy type, $E=\onehalf\dot{q}^\trans(\cof G)\dot{q}+W$ such that the eigenvalues of $\Gt^{-1}G$ are functionally independent.

A potential Newton system $\ddot{q}=-\grad V$ is of cofactor type with $\Gt=I$, and the accompanying first integral is the total energy $\tilde{E}=\onehalf\dot{q}^\trans\dot{q}+V$. The theorem
therefore says that it is separable if there is an extra first integral~$E$ of cofactor type with functionally independent eigenvalues. Its existence also implies the existence of a set of $n$ first integrals.

Benenti \cite{sb} has shown that the separability of a potential Newton system is equivalent to the existence of a very special first integral, corresponding to a ``characteristic'' Killing tensor $K$. For instance, in the case of systems separable in elliptic coordinates, this tensor is expressed through the ``inertia'' tensor $G=-qq^\trans+\diag(\lambda_1,\dots,\lambda_n)$ as $K=(\tr G)I-G$. It has also been noticed that the eigenvalues of $G$ are separation coordinates in this case.

In our terminology, $G$ is an elliptic coordinates matrix, and one can show that the eigenvalues of $G$ define elliptic coordinates (see \propositionref{genell}), which also justifies the terminology. Moreover, by the construction in the $2\Rightarrow n$ theorem \cite[Theorem~5.3]{hl}, $A^{(n-2)}=(\tr G)I-G$. When $n=2$, we thus have $K=A^{(0)}$ so that Benenti's characteristic first integral is the same as the first integral $E$ considered here. However, when $n>2$, the two types of first integrals are different, even though they both imply separability for the classical potential Newton system.
\end{remark}

\begin{remark}[Novel features]
By using generalized orthogonality \eqref{genort}, the transformation formula \eqref{momenta} for the momenta coordinates can be written as
\begin{displaymath}
s=(\Gt\Jac)^{-1}p.
\end{displaymath}
This can be compared with the corresponding formula for a canonical transformation of coordinates: a change of configuration coordinates $q\to u$ induces a change of momenta coordinates $p\to s$ according to 
\begin{displaymath}
s=\Jac^{-1}p.
\end{displaymath}
Thus we see that the change of coordinates considered in this paper is non-canonical, unless $\Gt=I$, which, as mentioned before, specializes the theory to the classical case of separable potential systems.

Another novel feature of this separability theorem is the fact that a related system, having the same trajectories, is solved instead of the original system. Again, if $\Gt=I$, the two systems coincide, and the theory specializes to the classical case.
\end{remark}

\begin{remark}[$\pmb{t\mapsto\tau}$]
\remarklabel{tt}
Solutions of the two systems \eqref{originalsystem} and \eqref{relatedsystem} are related by the function $\tau(t)$ defined as a solution to the differential equation
\begin{equation}
\label{tau}
\det\Gt(q\circ\tau)\cdot\frac{d\tau}{dt}=1,
\end{equation}
where $q$ is a solution to \eqref{relatedsystem}. Indeed, if $(q,p)$ solves \eqref{relatedsystem}, then $(q\circ\tau,p\circ\tau)$ solves \eqref{originalsystem} as is easily verified by the chain rule:
\begin{displaymath}
\frac{d}{dt}(q\circ\tau)=\Bigl(\frac{dq}{d\tau}\circ\tau\Bigr)\frac{d\tau}{dt}=
\det\Gt(q\circ\tau)\cdot(p\circ\tau)\cdot\frac{d\tau}{dt}=p\circ\tau,
\end{displaymath}
and similarly for $d(p\circ\tau)/dt=M(q\circ\tau)$. The solution of \eqref{tau} can be obtained by one further quadrature.

\end{remark}

\begin{remark}[$\pmb{2\Rightarrow n}$ theorem]
The quadratic first integrals referred to in \theoremref{cofseparability} coincide with those given by the $2\Rightarrow n$ theorem \cite[Theorem~5.3]{hl}, as follows from the relation
\begin{equation}
\label{kineticrel}
\Bigl(\frac{dq}{dt}\Bigr)^\trans\cof(G+\mu\Gt)\Bigl(\frac{dq}{dt}\Bigr)=
(\dG)
\sum_{k=0}^{n-1}\mu^k
\sum_{j=1}^n \sigmacheck{n-k-1}j \Delta_j s_j^2
\end{equation}
proved below. This relation shows that the first integrals considered in the two theorems have the same kinetic part and therefore have to be the same, since the Hamiltonian~\eqref{hamiltonian} has a non-trivial quadratic dependence on all momenta variables.

The separability theorem thus implies the $2\Rightarrow n$ theorem in the case of generic cofactor pair systems (since the separation coordinates $u$ are possible to define for such systems). See further \sectionref{functionalindep}.

\begin{proof}
 We have that $dq/dt=p$, if $t$ and $\tau$ are related as in \remarkref{tt}. Thus
 \begin{displaymath}
 \frac{dq}{dt}=\Jac^{-\trans}\Delta s.
 \end{displaymath}
 Further, the eigenvalue-eigenvector relation \eqref{eigrel} 
 gives $G\Jac=\Gt\Jac\mathcal{U}$, so that
 $\Jac^\trans G\Jac=\Delta\mathcal{U}$ by 
 generalized orthogonality \eqref{genort}. 
 It is now clear that (when $-\mu$ is not an eigenvalue of $X$)
 \begin{multline*}
        \Bigl(\frac{dq}{dt}\Bigr)^\trans\cof(G+\mu\Gt)\Bigl(\frac{dq}{dt}\Bigr)=
        \det(G+\mu\Gt)\,
        s^\trans\Delta\bigl(\Jac^\trans(G+\mu\Gt)\Jac\bigr)^{-1}\Delta s=\\
        (\dG)\det(X+\mu I)\,
        s^\trans\Delta(\mathcal{U}+\mu I)^{-1}s, 
 \end{multline*}
 which can be rewritten as a sum of products
 \begin{displaymath}
        (\dG)\sum_{j=1}^n\prod_{i=1}^n
        \frac{u_i+\mu}{(u_j+\mu)}\Delta_js_j^2. 
 \end{displaymath}
 By using \eqref{checksymmetricdefinition}, we can expand these products to
 find the right hand side of~\eqref{kineticrel}.
\end{proof}
\end{remark}

\begin{remark}[Hamiltonian separability for non-Hamiltonian systems]
\remarklabel{srwnonham}
The example of cofactor pair systems indicates how to give meaning to the concept of separability for general Newton systems $\ddot{q}=M(q)$ or more general second order dynamical systems. Any such system can be considered to be separable whenever there exist a transformation to new coordinates $(u,s)$ in which the transformed equations acquire the canonical Hamiltonian form and the related Hamilton--Jacobi equation can be solved through a separated ansatz $S=\sum S_k(u_k)$. Such separation can be even more general when the Hamiltonian for the transformed equations does not have the St\"ackel form as is necessarily the case for cofactor pair systems. 
\end{remark}

\begin{remark}[Separability of non-canonical Hamiltonian systems]
\remarklabel{srwnoncan}
The related system has a non-canonical Hamiltonian formulation in the $(2n+1)$-dimensional phase space with coordinates $(q,p,d)$, but it has no directly related Hamilton--Jacobi equation. The transformation $(q,p,d)\to(u,s,c)$ involving \eqref{sdef} given in~\cite{km-mb} is non-canonical and gives the Hamiltonian system a canonical form in $(2n+1)$-dimensional phase space, with a St\"ackel separable Hamiltonian. In this sense, a non-canonical Hamiltonian system can be considered separable. The important thing is, that a non-canonical Hamiltonian system has been turned (through a non-canonical transformation) into a canonical Hamiltonian system for which the concept of Hamilton--Jacobi equations and the concept of separability is well defined.
\end{remark}

\section{Proof of the separability theorem}

In this section, we prove \theoremref{cofseparability}. The proof is rather long and technical, but have been structured into subsections, to help the reader in keeping track of the logic. To simplify the notation, we now use a dot for $d/d\tau$.

\subsection{Hamiltonian formulation}\sectionlabel{hamiltonianform}

We begin by establishing a canonical Hamiltonian formulation of \eqref{relatedsystem}. First, we transform the equations using the transformation $(q,p)\to(u,s)$. Second, we show that the transformed equations can be expressed through the Hamiltonian~\eqref{hamiltonian}.

In order to transform the equation $\dot{q}=(\dG)p$, we use the chain rule in the form $\dot{q}=\Jac^{-\trans}\dot{u}$ and the definition of the momenta coordinates written as $p=\Jac^{-\trans}\Delta s$. When inserted, this gives immediately the equivalent form
\begin{displaymath}
\dot{u}=(\dG)\Delta s
\end{displaymath}
by canceling $\Jac^{-\trans}$. To give this equation a Hamiltonian formulation $\dot{u}=\partial_sH$, we need to find a function $H=H(u,s)$ such that $\partial_sH=(\dG)\Delta s$. This is an overdetermined system of PDEs, which can easily be integrated to give
\begin{displaymath}
H=\onehalf(\dG)\sum_{k=1}^n\Delta_ks_k^2+F(u),
\end{displaymath}
where $F$ is an arbitrary function of $u$.

We now turn to the equation $\dot{p}=(\dG)M$, and we use the representation $M=-(\cof\Gt)^{-1}\grad\Wt$ so that 
\begin{equation}
\label{peq}
\dot{p}=-\Gt\grad\Wt.
\end{equation}
The definition of the momenta coordinates gives $\partial_ps_i=\Delta_i^{-1}\grad u_i$, so that \eqref{peq} is equivalent to the system 
\begin{displaymath}
(\partial_p s_j)^\trans\dot{p}=
-\Delta_j^{-1}(\grad u_j)^\trans\Gt\grad\Wt,
\quad j=1,\dots,n.
\end{displaymath}
To get further, we need the following variant of the chain rule.

\begin{lemma}[Chain rule] 
 \lemmalabel{chainrule}
 \begin{displaymath}
        (\grad u_j)^\trans\Gt\grad=\Delta_j
        \frac{\partial}{\partial u_j}.
 \end{displaymath}
\end{lemma}

\begin{proof}
 The chain rule $\grad=\Jac\partial_u$ together with
 $\Gt\Jac=\Jac^{-\trans}\Delta$ from \eqref{genort} gives 
 \begin{displaymath}
        (\grad u_j)^\trans\Gt\grad=(\Jac^{-1}\grad u_j)^\trans
        \Delta\partial_u,
 \end{displaymath}
 which equals~$\Delta_j\,\partial/\partial u_j$ since $\Jac^{-1}\grad u_j$ forms
 the $j$:th standard basis vector.
\end{proof}

From the usual chain rule we have

\begin{displaymath}
\dot{s}_j=(\grad s_j)^\trans\dot{q}+(\partial_p s_j)^\trans\dot{p},
\end{displaymath}
which together with \lemmaref{chainrule} shows that \eqref{peq} is equivalent to
\begin{displaymath}
\dot{s}_j=(\grad s_j)^\trans\dot{q}-\frac{\partial\Wt}{\partial u_j},
\quad j=1,\dots,n.
\end{displaymath}
These equations can be given a Hamiltonian formulation $\dot{s}=-\partial_uH$ with the same Hamiltonian as above, if $(\grad s_j)^\trans\dot{q}=-(\partial/\partial u_j)(H-F)$ and $F=\Wt$. The latter condition can be taken as a definition since $F$ is arbitrary, but it remains to show the former condition. This is more complicated, and will be carried out in the next section.

\subsection{Kinetic part}\sectionlabel{kineticpart}
 
We shall show that
\begin{equation}
\label{kinpart}
        (\grad s_j)^\trans\dot{q}=
        -\frac{\partial}{\partial u_j}\biggl(
        \onehalf(\dG)\sum_{k=1}^n\Delta_ks_k^2\biggr).
\end{equation}
This will be done using $\dot{q}=(\dG)p$, which can be written as 
\begin{displaymath}
\dot{q}=(\dG)\Gt\Jac s,
\end{displaymath}
since it is possible to express $p=\Jac^{-\trans}\Delta s$ through 
\begin{displaymath}
p=\Gt\Jac s
\end{displaymath}
as follows from the generalized orthogonality \eqref{genort}.

We expand
\begin{equation}
\label{kinlhs}
 (\grad s_j)^\trans\dot{q}=(\dG)\sum_{a,e,f}
 \frac{\partial s_j}{\partial q_a}
 \Gt_{ae}
 \frac{\partial u_f}{\partial q_e}
 s_f.
\end{equation}
The derivatives of $s_j$ can be calculated from $s=\Delta^{-1}\Jac^\trans p$ as
\begin{displaymath}
 \frac{\partial s_j}{\partial q_a}=
 \sum_{b,c,d}\frac{\partial}{\partial q_a}\Bigl(
 \Delta_j^{-1}\frac{\partial u_j}{\partial q_b}\Bigr)
 \Gt_{bd}\frac{\partial u_c}{\partial q_d}s_c,
\end{displaymath}
where the expression $\Gt\Jac s$ has been substituted for $p$ after differentiation. To simplify this, we use Leibniz' rule and the generalized orthogonality \eqref{genort} to find
\begin{equation}
\label{dsj}
 \frac{\partial s_j}{\partial q_a}=
 \frac{\partial \Delta_j^{-1}}{\partial q_a}
 \Delta_j s_j + \Delta_j^{-1}
 \sum_c\Bigl(\grad\frac{\partial u_j}{\partial q_a}\Bigr)^\trans
 \Gt\grad u_c \,s_c.
\end{equation}
The Hessian of $u_j$ appears here; we denote it by
\begin{displaymath}
 H_j=\grad(\grad u_j)^\trans
\end{displaymath}
in the subsequent. By combining \eqref{kinlhs} and \eqref{dsj}, we find
\begin{multline*}
 (\grad s_j)^\trans\dot{q}=
 (\dG)\Delta_j\sum_f(\grad\Delta_j^{-1})^\trans
 \Gt\grad u_f
 \,s_f \,s_j+\\
 (\dG)\Delta_j^{-1}\sum_{c,f}(\grad u_f)^\trans
 \Gt H_j \Gt \grad u_c
 \,s_f \,s_c,
\end{multline*}
which we rewrite as 
\begin{equation}
\begin{split}
\label{dsjfinal}
 (\grad s_j)^\trans\dot{q}&=(\dG)\times\\
 \biggl(
 &\vphantom{\sum_{k\ne j}}
 \bigl[\Delta_j(\grad\Delta_j^{-1})^\trans\Gt\grad u_j+
 \Delta_j^{-1}(\grad u_j)^\trans\Gt H_j\Gt \grad u_j\bigr]s_j^2+\\
 &\sum_{k\ne j}\bigl[\Delta_j^{-1}
 (\grad u_k)^\trans\Gt H_j\Gt \grad u_k\bigr]s_k^2+\\
 &\sum_{k\ne j}\bigl[
 \Delta_j(\grad\Delta_j^{-1})^\trans\Gt\grad u_k+
 2\Delta_j^{-1}(\grad u_j)^\trans\Gt H_j\Gt \grad u_k
 \bigr] s_k \,s_j+\\
 &\sum_{\makebox[0pt]{$\scriptstyle k,\ell,j\ne$}}\bigl[
 \Delta_j^{-1}(\grad u_k)^\trans\Gt H_j\Gt \grad u_\ell\bigr]s_k \,s_\ell
 \biggr)
\end{split}
\end{equation}
to single out the coefficients of different products of $s_k$.

In the next four propositions, we show that the coefficients of mixed
products vanish, and that the other coefficients are of the form
\begin{displaymath}
 -\onehalf\frac{\partial}{\partial u_j}\bigl(
 (\dG)\Delta_k\bigr).
\end{displaymath}

For the proofs of these propositions, we need an elegant representation formula for the gradients of the generalized metric coefficients. This formula also yields useful identities when extended to the ``off-diagonal'' coefficients.

\begin{lemma}[Representation formula]
 \lemmalabel{representation}
 The gradient of $(\grad u_j)^\trans\Gt\grad u_k$ 
 can be represented as
 \begin{displaymath}
        \grad(\delta_{jk}\Delta_j)=
        H_j\Gt\grad u_k+H_k\Gt\grad u_j+
        \bigl((\grad u_j)^\trans\Nt\bigr)\grad u_k+
        \bigl((\grad u_k)^\trans\Nt\bigr)\grad u_j.
 \end{displaymath}
 Thus
 \begin{displaymath}
        \grad\Delta_j=
        2H_j\Gt\grad u_j+
        2\bigl((\grad u_j)^\trans\Nt\bigr)\grad u_j
 \end{displaymath}
 and
 \begin{displaymath}
        H_j\Gt\grad u_k+H_k\Gt\grad u_j+
        \bigl((\grad u_j)^\trans\Nt\bigr)\grad u_k+
        \bigl((\grad u_k)^\trans\Nt\bigr)\grad u_j=0,\quad j\ne k.
 \end{displaymath}
\end{lemma}

\begin{proof}
 By Leibniz' rule, 
 \begin{displaymath}
        \grad\sum_{a,b}\Gt_{ab}
        \frac{\partial u_j}{\partial q_a}
        \frac{\partial u_k}{\partial q_b}=
        \sum_{a,b}\grad\Gt_{ab}
        \frac{\partial u_j}{\partial q_a}
        \frac{\partial u_k}{\partial q_b}+
        H_j\Gt\grad u_k+H_k\Gt\grad u_j.
 \end{displaymath}
 The relation $\partial\Gt_{ab}/\partial q_\ell=
 \delta_{a\ell}\Nt_b+\delta_{b\ell}\Nt_a$ now gives the desired 
 representation formula, from which the other two follow immediately.
\end{proof}

\begin{proposition}
 For all $j=1,\dots,n$,
 \begin{displaymath}
        \Delta_j(\grad\Delta_j^{-1})^\trans\Gt\grad u_j+
        \Delta_j^{-1}(\grad u_j)^\trans\Gt H_j\Gt \grad u_j=
        -\onehalf\Delta_j\frac{\partial}{\partial u_j} 
        \log\bigl((\dG)\Delta_j\bigr).
 \end{displaymath}
\end{proposition}

\begin{proof}
 By writing the left hand side as
 \begin{displaymath}
        -\onehalf(\grad u_j)^\trans\Gt\bigl(
        2\grad\log\Delta_j-
        2\Delta_j^{-1}H_j\Gt\grad u_j\bigr),
 \end{displaymath}
 we can use \lemmaref{representation} and \eqref{assvecrepr}
 to further rewrite it as
 \begin{multline*}
        -\onehalf(\grad u_j)^\trans\Gt\bigl(
        \grad\log\Delta_j+
        2\Delta_j^{-1}(\grad u_j)^\trans(\onehalf\Gt\grad\log\dG)
        \grad u_j\bigr)=\\
        -\onehalf(\grad u_j)^\trans\Gt\grad\bigl(
        \log\Delta_j+
        \log\dG\bigr),
 \end{multline*}
 which, by \lemmaref{chainrule}, equals the 
 right hand side in the proposition.
\end{proof}

\begin{proposition}
 For all distinct $j,k=1,\dots,n$,
 \begin{displaymath}
        \Delta_j^{-1}(\grad u_k)^\trans\Gt H_j\Gt \grad u_k=
        -\onehalf\Delta_k\frac{\partial}{\partial u_j} 
        \log\bigl((\dG)\Delta_k\bigr).
 \end{displaymath}
\end{proposition}

\begin{proof}
 \lemmaref{representation} and the generalized orthogonality \eqref{genort}
 shows that 
 the left hand side equals
 \begin{multline*}
        -\Delta_j^{-1}(\grad u_k)^\trans\Gt\bigl[ 
        H_k\Gt\grad u_j+
        \bigl((\grad u_j)^\trans\Nt\bigr)\grad u_k+
        \bigl((\grad u_k)^\trans\Nt\bigr)\grad u_j\bigl]=\\
        -\Delta_j^{-1}\bigl[ 
        (\grad u_k)^\trans\Gt H_k\Gt\grad u_j+\Delta_k
        (\grad u_j)^\trans\Nt\bigl].
 \end{multline*}
 By using \propositionref{beta}, we can further rewrite it as
 \begin{multline*}
        -\onehalf\Delta_j^{-1}\Delta_k\bigl[ 
        (\grad\log\Delta_k)^\trans\Gt\grad u_j+2(\grad u_j)^\trans\Nt\bigl]=\\
        -\onehalf\Delta_j^{-1}\Delta_k(\grad u_j)^\trans\bigl[ 
        \Gt\grad\log\Delta_k+2\Nt\bigl],
 \end{multline*}
 which, by \eqref{assvecrepr} and \lemmaref{chainrule}, 
 equals the right hand side in the proposition.
\end{proof}

\begin{proposition}
 \propositionlabel{beta}
 For all distinct $j,k=1,\dots,n$,
 \begin{displaymath}
        \Delta_j(\grad\Delta_j^{-1})^\trans\Gt\grad u_k+
        2\Delta_j^{-1}(\grad u_j)^\trans\Gt H_j\Gt \grad u_k=0.
 \end{displaymath}
\end{proposition}

\begin{proof}
 The left hand side can be rewritten
 by applying \lemmaref{representation} to the factor
 $\Delta_j(\grad\Delta_j^{-1})^\trans=
 -\Delta_j^{-1}(\grad\Delta_j)^\trans$, after which the
 result will vanish because of the generalized orthogonality \eqref{genort}.
\end{proof}

\begin{proposition}
 For all distinct $j,k,\ell=1,\dots,n$,
 \begin{displaymath}
        (\grad u_k)^\trans\Gt H_j\Gt \grad u_\ell=0.
 \end{displaymath}
\end{proposition}

\begin{proof}
 \lemmaref{representation} states that
 \begin{displaymath}
        H_j\Gt\grad u_k+H_k\Gt\grad u_j+
        \bigl((\grad u_j)^\trans\Nt\bigr)\grad u_k+
        \bigl((\grad u_k)^\trans\Nt\bigr)\grad u_j=0.
 \end{displaymath}
 If this is multiplied from the left by $(\grad u_\ell)^\trans\Gt$,
 we have, due to the generalized orthogonality \eqref{genort},
 \begin{displaymath}
        (\grad u_\ell)^\trans\Gt H_j\Gt\grad u_k+
        (\grad u_\ell)^\trans\Gt H_k\Gt\grad u_j=0.
 \end{displaymath}
 By repeating this process for all other combinations of indices,
 we get two more equations of this form. Together with three
 trivial symmetry relations 
 \begin{displaymath}
        (\grad u_\ell)^\trans\Gt H_j\Gt\grad u_k-
        (\grad u_k)^\trans\Gt H_j\Gt\grad u_\ell=0
 \end{displaymath}
 that arise by transposition of the involved scalars, 
 we get in total six homogeneous
 algebraic equations for six unknowns of type 
 $(\grad u_k)^\trans\Gt H_j\Gt \grad u_\ell$. 
 This system is uniquely solvable, so the only solution is
 the zero solution. 
\end{proof}

The preceding four propositions together with \eqref{dsjfinal} finally completes the proof of \eqref{kinpart}.

\subsection{St\"ackel form}\sectionlabel{stackelform}

In the following two propositions we show that the Hamiltonian \eqref{hamiltonian} has the St\"ackel form
\begin{displaymath}
        H=\sum_{k=1}^n\frac{1}{U'(u_k)}
        \bigr(\onehalf f_k(u_k)\,s_k^2+g_k(u_k)\bigl),
\end{displaymath}
where $f_k$ and $g_k$ are some functions of one variable only.

\begin{proposition}
 \propositionlabel{stackelkinetic}
 \begin{displaymath}
        (\dG)\Delta_k=\frac{f_k(u_k)}{U'(u_k)}
 \end{displaymath}
 for some functions $f_k$ of one variable.
\end{proposition} 

The proof follow the proof of Proposition 32 in \cite{hl-srw}.

\begin{proof}
 We will prove that
 $\grad[(\dG)U'(u_j)\Delta_j]=\lambda_j\grad u_j$ for some function
 $\lambda_j$. This identity implies the proposition, for
 if it is multiplied by $(\grad u_k)^\trans\Gt$
 from the left, it follows from \lemmaref{chainrule} together
 with the generalized orthogonality \eqref{genort} that
 $(\dG)U'(u_j)\Delta_j$ only depends on $u_j$.

 In order to establish the above identity, let us calculate
 the gradient of the identity $\det(G-u_j\Gt)\equiv0$; we get
 \begin{equation}
        \label{deltaid}
        \frac{\partial}{\partial\mu}\det(G-\mu\Gt)\grad u_j+
        \grad\det(G-\mu\Gt)\equiv0,\quad\text{when $\mu=u_j.$}
 \end{equation} 
 For the first term we have immediately
 \begin{displaymath}
        \frac{\partial}{\partial\mu}\det(G-\mu\Gt)=
        (-1)^n(\dG)U'(u_j),\quad\text{when $\mu=u_j.$}
 \end{displaymath} 
 For the second term we use \eqref{assvecrepr}
 applied with the pencil $N-\mu\Nt$ to get
 \begin{equation}
        \label{deltasecond}
        \grad\det(G-\mu\Gt)=
        2\cof(G-\mu\Gt)(N-\mu\Nt).
 \end{equation}
 Next, we multiply \eqref{deltaid} by $2(N^\trans-\Nt^\trans X)$ 
 from the left to find
 \begin{displaymath}
        (-1)^n(\dG)U'(u_j)\Delta_j+
        4(N-u_j\Nt)^\trans\cof(G-u_j\Gt)(N-u_j\Nt)=0
 \end{displaymath} 
 (we postpone the technical details to the end of the present
 proof). We have now control over the gradient of
 $(\dG)U'(u_j)\Delta_j$; it is proportional to
 \begin{multline*}
        \grad[(N-u_j\Nt)^\trans\cof(G-u_j\Gt)(N-u_j\Nt)]=\\
        \frac{\partial}{\partial\mu}
        [(N-\mu\Nt)^\trans\cof(G-\mu\Gt)(N-\mu\Nt)]
        \grad u_j+\\
        \grad[(N-\mu\Nt)^\trans\cof(G-\mu\Gt)(N-\mu\Nt)],\quad
        \text{when $\mu=u_j.$}
 \end{multline*} 
 In fact, the last term here is also proportional to $\grad u_j$,
 since
 \begin{displaymath}
        \grad[(N-\mu\Nt)^\trans\cof(G-\mu\Gt)(N-\mu\Nt)]=
        2(\alpha-\mu\tilde{\alpha})\cof(G-\mu\Gt)(N-\mu\Nt)
 \end{displaymath} 
 and 
 \begin{displaymath}
        2\cof(G-\mu\Gt)(N-\mu\Nt)=
        \frac{\partial}{\partial\mu}\det(G-\mu\Gt)\grad u_j,
        \quad\text{when $\mu=u_j,$}
 \end{displaymath}
 as follows from \eqref{deltasecond} and \eqref{deltaid}.
 
 To complete the proof, we need to show two identities. The first
 one is $2(N^\trans-\Nt^\trans X)\grad u_j=\Delta_j$.
 From \eqref{assvecrepr}
 applied with both $N$ and $\Nt$ and from \eqref{xid} 
 we find
 \begin{multline*}
        2(N^\trans-\Nt^\trans X)\grad u_j=
        [(\grad\log\det G)^\trans G-
        (\grad\log\det\Gt)^\trans\Gt X]\grad u_j=\\
        (\grad\log\det X)^\trans G\grad u_j=
        (\grad\tr X)^\trans X^{-\trans}G\grad u_j=
        (\grad u_j)^\trans\Gt\grad\tr X,        
 \end{multline*} 
 which by \lemmaref{chainrule} equals $\Delta_j$.
 
 The second identity we need to show is
 \begin{multline*}
        (N^\trans-\Nt^\trans X)\cof(G-u_j\Gt)(N-u_j\Nt)=\\
        (N-u_j\Nt)^\trans\cof(G-u_j\Gt)(N-u_j\Nt),
 \end{multline*}
 which is immediate once we have observed that $(X-\mu
 I)\cof(G-\mu\Gt)= \det(X-\mu I)\cof\Gt$ vanishes when $\mu=u_j$. 
 Indeed,
 \begin{multline*}
        (N^\trans-\Nt^\trans X)\cof(G-\mu\Gt)(N-\mu\Nt)=\\
        (N-\mu\Nt)^\trans\cof(G-\mu\Gt)(N-\mu\Nt)-
        \Nt^\trans(X-\mu I)\cof(G-\mu\Gt)(N-\mu\Nt),
 \end{multline*}
 which in turn equals 
 $(N-\mu\Nt)^\trans\cof(G-\mu\Gt)(N-\mu\Nt)$ when $\mu=u_j$.
\end{proof}

\begin{proposition}
 \propositionlabel{stackelpotential}
 \begin{equation}
        \Wt=\sum_{i=1}^n\frac{1}{U'(u_i)}g_i(u_i),
 \end{equation}
 for some functions $g_i$ of one variable. 
\end{proposition}

\begin{proof}
 The idea is to solve the Frobenius compatibility conditions for
 Equation~\eqref{prefe} satisfied by any pair of
 quasi-potentials belonging to the same cofactor pair system. In terms
 of $u$ coordinates, this equation can be written as
 $X\Jac\partial_u W = (\det X)\Jac\partial_u\Wt$, or
 \begin{displaymath}
 \mathcal{U}\partial_u W = (\det\mathcal{U})\partial_u\Wt
 \end{displaymath}
 with $\mathcal{U}=\Jac^{-1}X\Jac$ as in \eqref{eigrel}. 
 Since $\mathcal{U}$ is diagonal, we find
 \begin{displaymath}
        \frac{\partial W}{\partial u_k} = 
        \biggl(\prod_{i\ne k} u_i\biggr)
        \frac{\partial \Wt}{\partial u_k},\quad k=1,\dots,n.
 \end{displaymath}
 For $W$ to exist (we assume that it is in $C^\infty$), we
 require its mixed second order partial derivatives with respect to
 $k$ and $\ell$ to be equal. This gives
 \begin{displaymath}
        \frac{\partial \Wt}{\partial u_k}+
        u_\ell\frac{\partial^2 \Wt}{\partial u_k\partial u_\ell}=
        \frac{\partial \Wt}{\partial u_\ell}+
        u_k\frac{\partial^2 \Wt}{\partial u_\ell\partial u_k},
        \quad k,\ell=1,\dots,n.
 \end{displaymath}
These equations can be written in the form $\partial^2(u_k-u_\ell)\Wt/\partial u_k\partial u_\ell=0$, which according to \corollaryref{annoying-system} have the indicated solution.
\end{proof}

\subsection{Quadratic first integrals}\sectionlabel{qfis}

We shall now show that the functions
\begin{displaymath}
        E^{(k)}=\sum_{i=1}^n\frac{\sigmacheck{n-k-1}i}{U'(u_i)}
        \bigl(\onehalf f_i(u_i)\,s_i^2+g_i(u_i)\bigr), \quad k=0,\dots,n-1,
\end{displaymath} 
with $f_i$ from \propositionref{stackelkinetic} and $g_i$ from \propositionref{stackelpotential}, are first integrals for the Hamiltonian system governed by the Hamiltonian~\eqref{hamiltonian}, that is, $H=E^{(n-1)}$. This fact can be expressed in terms of the Poisson bracket as $\{H,E^{(k)}\}=0$. Since $H$ has the same form as all other functions $E^{(k)}$, it requires no extra effort to show that $\{E^{(k)},E^{(\ell)}\}=0$ for all $k$, $\ell$. This means precisely that all first integrals $E^{(k)}$ are in involution.

\begin{proposition} 
 \propositionlabel{involution}
 The functions $E^{(k)}$ are first integrals in involution.
\end{proposition}

\begin{proof}
A straightforward calculation shows that
\begin{multline*}       
        \{E^{(n-k-1)},E^{(n-\ell-1)}\}=
        \sum_{a=1}^n\frac{f_a(u_a)\,s_a}{U'(u_a)}\times\\
        \sum_{i=1}^n\biggl\{
        \onehalf s_i^2
        \left[
        \frac{\partial}{\partial u_a}
        \left(\frac{\sigmacheck{k}{i}}{U'(u_i)}f_i(u_i)\right)\sigmacheck{\ell}{a}
        -
        \frac{\partial}{\partial u_a}
        \left(\frac{\sigmacheck{\ell}{i}}{U'(u_i)}f_i(u_i)\right)\sigmacheck{k}{a}
        \right]+\\
        \left[
        \frac{\partial}{\partial u_a}
        \left(\frac{\sigmacheck{k}{i}}{U'(u_i)}g_i(u_i)\right)\sigmacheck{\ell}{a}
        -
        \frac{\partial}{\partial u_a}
        \left(\frac{\sigmacheck{\ell}{i}}{U'(u_i)}g_i(u_i)\right)\sigmacheck{k}{a}
        \right]
        \biggr\}.
\end{multline*}
This is identically zero since the expressions within square brackets
vanish. The only property of the functions $f_i$ and $g_i$ needed to
see this is that they only depend on one variable each. 

If $a=i$, we
have $\partial\sigmacheck{k}{i}/\partial u_a =\partial\sigmacheck{\ell}{i}/\partial
u_a=0$, so that it is possible to take these
symmetric polynomials outside the respective derivative, showing that the two
terms cancel. 

If $a\ne i$, we can factor out $f_i(u_i)$
or $g_i(u_i)$ and thus have to show that 
\begin{displaymath}
        \frac{\partial}{\partial u_a}
        \left(\frac{\sigmacheck{k}{i}}{U'(u_i)}\right)\sigmacheck{\ell}{a}
        -
        \frac{\partial}{\partial u_a}
        \left(\frac{\sigmacheck{\ell}{i}}{U'(u_i)}\right)\sigmacheck{k}{a}
        =0.
\end{displaymath}
Since 
$\partial\bigl(1/U'(u_i)\bigr)/\partial u_a=1/\bigl(U'(u_i)(u_i-u_a)\bigr)$, it is equivalent to show that
\begin{multline*}
        (u_i-u_a)\frac{\partial}{\partial u_a}\sigmacheck{k}{i}\,
        \sigmacheck{\ell}{a}+\sigmacheck{k}{i}\,
        \sigmacheck{\ell}{a}-\\
        (u_i-u_a)\frac{\partial}{\partial u_a}\sigmacheck{\ell}{i}\,
        \sigmacheck{k}{a}-\sigmacheck{\ell}{i}\,
        \sigmacheck{k}{a}=0.
\end{multline*}
From \eqref{derivativeofsymmetric} and \eqref{decomposesymmetric} follows that the left hand side can be written as
\begin{multline*}
(u_i-u_a)\sigma_{k-1}(u_i\sigma_{\ell-1}+\sigma_\ell)+
(u_a\sigma_{k-1}+\sigma_k)(u_i\sigma_{\ell-1}+\sigma_\ell)-\\
(u_i-u_a)\sigma_{\ell-1}(u_i\sigma_{k-1}+\sigma_k)-
(u_a\sigma_{\ell-1}+\sigma_\ell)(u_i\sigma_{k-1}+\sigma_k)
\end{multline*}
in terms of elementary symmetric polynomials depending on all $u_1,\dots,u_n$ except $u_i$ and $u_a$. This expression is easily seen to vanish.
\end{proof}

The first integrals $E^{(k)}$ are functionally independent as functions on $\R^{2n}=\{(u,s)\}$. The standard test for this is that the differentials $dE^{(k)}$ should be linearly independent. Here it is simple to verify this fact, since the matrix $(\partial E^{(k)}/\partial s_\ell)$ can be written as a product of two non-singular matrices: the inverse Vandermonde matrix \eqref{vandermondeinverse} and $\diag(f_1,\dots,f_n)$.

\subsection{Solution by quadratures}\sectionlabel{quadrature}

We shall now use our previous results to obtain the solution of the cofactor pair system \eqref{originalsystem} by quadratures.

We begin by noting that the first integrals separate in the following sense. Denote by $\alpha_k$ the constant value of $E^{(k)}$, and consider the equations
\begin{equation}
\label{hjeqns}
E^{(k)}(u,s)=\alpha_k,\quad k=0,\dots,n-1.
\end{equation}
These equations can be put into matrix form as
        \begin{displaymath}
         \begin{pmatrix}
        \displaystyle{\frac{\sigmacheck01}{U'(u_1)}} &
        \dotsm &
        \displaystyle{\frac{\sigmacheck0n}{U'(u_n)}} \\
        \hdotsfor{3} \\
        \displaystyle{\frac{\sigmacheck{n-1}1}{U'(u_1)}} & 
    \dotsm &
        \displaystyle{\frac{\sigmacheck{n-1}n}{U'(u_n)}}
         \end{pmatrix}
         \begin{pmatrix}
                \onehalf f_1(u_1)\,s_1^2+g_1(u_1) \\
                \hdotsfor{1} \\
                \onehalf f_n(u_n)\,s_n^2+g_n(u_n) 
         \end{pmatrix}
         =
         \begin{pmatrix}
                \alpha_{n-1} \\
                \hdotsfor{1} \\
                \alpha_{0} \\
         \end{pmatrix},
        \end{displaymath}
from which it is clear that they are equivalent to
        \begin{displaymath}
         \begin{pmatrix}
                \onehalf f_1(u_1)\,s_1^2+g_1(u_1) \\
                \hdotsfor{1} \\
                \onehalf f_n(u_n)\,s_n^2+g_n(u_n) 
         \end{pmatrix}
         = (-1)^{n+1}
         \begin{pmatrix}
                (-u_1)^{n-1} &
                \dots &
                (-u_1)^0 \\
                \hdotsfor{3} \\
                (-u_n)^{n-1} &
                \dots &
                (-u_n)^0 \\
         \end{pmatrix}
         \begin{pmatrix}
                \alpha_{n-1} \\
                \hdotsfor{1} \\
                \alpha_{0} \\
         \end{pmatrix}.
        \end{displaymath}
Hence \eqref{hjeqns} is equivalent to the system
\begin{equation}
         \onehalf f_k(u_k)\,s_k^2+g_k(u_k)= 
         (-1)^{n+1}\sum_{i=0}^{n-1}(-u_k)^i\alpha_i,\quad k=1,\dots,n,
\end{equation}
consisting of relations involving only the $k$:th coordinate pair $(u_k,s_k)$.

We then define a function $S(u)=\sum S_k(u_k)$ by requiring the functions $S_k$ to be solutions to their respective ``separation equations''
\begin{displaymath}
        \onehalf f_k(u_k)
        \biggl(\frac{dS_k}{du_k}\biggr)^2
        +g_k(u_k)= 
        (-1)^{n+1}\sum_{i=0}^{n-1}(-u_k)^i\alpha_i.
\end{displaymath}
The function $S$ so constructed will thus be a simultaneous separated solution to the Hamilton--Jacobi equations
\begin{displaymath}
E^{(k)}(u,\partial_uS)=\alpha_k,\quad k=0,\dots,n-1.
\end{displaymath}
In particular this holds for the Hamiltonian $H=E^{(n-1)}$, that is, \eqref{hamiltonian}. By means of the classical Hamilton--Jacobi method we can therefore obtain the solution $(u(\tau),s(\tau))$ of the Hamiltonian system corresponding to $H$ by quadratures. By transforming this solution to $(q,p)$ variables, we also have the solution $(q(\tau),p(\tau))$ of the system \eqref{relatedsystem}. 

Finally, to get the solution of the cofactor pair system \eqref{originalsystem}, we need to find the function $\tau(t)$ from the differential equation \eqref{tau}, which requires one further quadrature. The solution for the cofactor pair system can then be obtained as $(q(\tau(t)),p(\tau(t)))$.

\subsection{Conclusion}

In \sectionref{hamiltonianform} and \sectionref{kineticpart} we established a canonical Hamiltonian form for the system \eqref{relatedsystem} related to the cofactor pair system \eqref{originalsystem}. In \sectionref{stackelform} we showed that the corresponding Hamiltonian has St\"ackel form, which made it possible to construct $n$ quadratic first integrals in \sectionref{qfis}. In \sectionref{quadrature} we used the St\"ackel form to separate variables in the corresponding Hamilton--Jacobi equation, providing the solution to the related system. The solution to the cofactor pair system was then obtained by rescaling the time.

\medskip\noindent
This completes the proof of \theoremref{cofseparability}.

\section{Separation coordinates for generic cofactor pair systems}

In the separability theorem we defined the separation coordinates $u_k$ for the cofactor pair system $d^2q/dt^2=-(\cof G)^{-1}\grad W=-(\cof\Gt)^{-1}\grad\Wt$ as roots of the polynomial $\det(G-\mu\Gt)$, provided that the functions $u_k(q)$ are functionally independent. Equivalently it can be said that $u_k$ are eigenvalues of $X=\Gt^{-1}G$, since $\dG\ne0$. 

In order to better explain the nature of these separation coordinates, we shall derive explicit formulas for the coordinate surfaces, which are ``non-confocal quadrics''. These formulas show how these new separation coordinates relate to the well known elliptic and parabolic coordinates that frequently occur in the classical separability theory.

\subsection{Classical elliptic and parabolic coordinates}

When $\Gt=I$, the cofactor pair system can be thought of as a classical potential Newton system $d^2q/dt^2=-\grad\Wt$ admitting an extra quadratic first integral $E=\onehalf\dot{q}^\trans(\cof G)\dot{q}+W$ in addition to the energy integral. In the generic case, the separation coordinates $u_k$ are then eigenvalues of $G$, which define elliptic or parabolic coordinates as explained by the following three propositions \cite{hl}. 

The first proposition shows that the matrix $G$ can be given a particularly simple standard form by a change of Euclidean reference frame $q\to Sq+v$ (such an affine transformation is called Euclidean if $S$ is an $n\times n$ orthogonal matrix, $S^\trans S=I$, with $\det S=1$, and $v\in\R^n$ is a translation vector). Under this transformation the quadratic forms 
\begin{displaymath}
\dot{q}^\trans \cof[G(q)] \dot{q}
\quad\text{and}\quad
\dot{q}^\trans \dot{q}
\end{displaymath}
go into
\begin{displaymath}
\dot{q}^\trans \cof[S^\trans G(Sq+v)S] \dot{q}
\quad\text{and}\quad
\dot{q}^\trans \dot{q}
\end{displaymath}
so that we still have a potential Newton system with an extra quadratic first integral. One can then give $S^\trans G(Sq+v)S$ one of the stated standard forms by a suitable choice of $S$ and $v$, as in the proofs of \theoremref{CPSell} and \theoremref{CPSpar} below.

\begin{proposition}[Standard form]
Let $G(q)=\alpha qq^\trans+\beta q^\trans+q\beta^\trans+\gamma$, with $\alpha\in\R$ and $\beta\in\R^n$ not both zero, be an elliptic coordinates matrix. There is a Euclidean change of reference frame $q\to Sq+v$, which gives $G$ one of the following standard forms: if $\alpha\ne0$ then
\begin{displaymath}
G(q)=-qq^\trans+\diag(\lambda_1,\dots,\lambda_n),
\end{displaymath}
and if $\alpha=0$ but $\beta\ne0$ then
\begin{displaymath}
G(q)=e_nq^\trans+qe_n^\trans+\diag(\lambda_1,\dots,\lambda_{n-1},0),
\quad e_n=(0,\dots,0,1)^\trans.
\end{displaymath}
\end{proposition}

The following two propositions can easily be proved by using a Weinstein--Aronszajn formula, as in the proofs of \theoremref{CPSell} and \theoremref{CPSpar} below. These propositions justify the name elliptic coordinates matrix for $G$.

\begin{proposition}[Elliptic coordinates]
\propositionlabel{genell}
If 
\begin{displaymath}
G(q)=-qq^\trans+\diag(\lambda_1,\dots,\lambda_n),
\end{displaymath}
then the eigenvalues $u_1(q)$, \dots, $u_n(q)$ of $G$ satisfy the rational equation
\begin{equation}
\label{ell}
1+\sum_{i=1}^n\frac{q_i^2}{\mu-\lambda_i}=
\prod_{j=1}^n(\mu-u_j)\bigg/
\prod_{k=1}^n(\mu-\lambda_k),
\end{equation}
which, if $\lambda_1<\dots<\lambda_n$, is the defining equation for elliptic coordinates~$u_j$ with parameters $\lambda_k$.
\end{proposition}

\begin{proposition}[Parabolic coordinates]
\propositionlabel{genpar}
If 
\begin{displaymath}
G(q)=e_nq^\trans+qe_n^\trans+\diag(\lambda_1,\dots,\lambda_{n-1},0),
\quad e_n=(0,\dots,0,1)^\trans,
\end{displaymath}
then the eigenvalues $u_1(q)$, \dots, $u_n(q)$ of $G$ satisfy the rational equation
\begin{equation}
\label{par}
2q_n-\mu+\sum_{i=1}^{n-1}\frac{q_i^2}{\mu-\lambda_i}=
-\prod_{j=1}^n(\mu-u_j)\bigg/
\prod_{k=1}^{n-1}(\mu-\lambda_k),
\end{equation}
which, if $\lambda_1<\dots<\lambda_{n-1}$, is the defining equation for parabolic coordinates~$u_j$ with parameters $\lambda_k$.
\end{proposition}

Equation \eqref{ell} define coordinate functions $u_k(q)$ as zeros of the rational function (of $\mu$) in the left hand side. It is easy to see that this function has $n$ distinct real roots $u_1(q)$, \dots, $u_n(q)$ such that
\begin{displaymath}
u_1<\lambda_1<u_2<\lambda_2<\dots<u_n<\lambda_n
\end{displaymath}
for each $n$-tuple $(q_1,\dots,q_n)$ of non-zero reals $q_k$. All these roots $u_k$ clearly satisfy
\begin{displaymath}
\sum_{i=1}^n\frac{q_i^2}{\lambda_i-u}=1,
\end{displaymath}
from which it follows that the hypersurfaces $u(q)=c$ of constant value of $u(q)$ are confocal quadrics (quadratic surfaces) in $\R^n$. Their character is decided by the value of $c$: for $c<\lambda_1$, the equation determines a family of ellipsoids, for $\lambda_1<c<\lambda_2$, a family of one-sheeted hyperboloids, and so on, over various families of hyperboloids, until $\lambda_{n-1}<c<\lambda_n$ (when $c>\lambda_n$, there are no real solutions for $q$). The functions $u_k(q)$ are then called elliptic coordinates. 

Similarly, Equation \eqref{par} define parabolic coordinates $u_k(q)$ satisfying
\begin{displaymath}
u_1<\lambda_1<u_2<\lambda_2<\dots<\lambda_{n-1}<u_n
\end{displaymath}
and
\begin{displaymath}
u+\sum_{i=1}^{n-1}\frac{q_i^2}{\lambda_i-u}=2q_n,
\end{displaymath}
which also define a family of confocal quadrics.

In both cases, the coordinate surfaces are orthogonal to each other as follows from the generalized orthogonality condition $(\grad u_j)^\trans\Gt\grad u_k=0$ with $\Gt=I$. That is, the elliptic and parabolic coordinate systems are orthogonal coordinate systems. 

In general, for arbitrary cofactor pair systems $\Gt\ne I$, and the separation coordinates are non-orthogonal with respect to the standard scalar product. However, they are always orthogonal with respect to~$\Gt(q)$.

\subsection{Cofactor-elliptic coordinates}

We shall now derive rational equations similar to \eqref{ell} and \eqref{par} for coordinates defined by two arbitrary elliptic coordinates matrices $G$ and $\Gt$. To this end, we shall use a formula for the Weinstein--Aronszajn determinant \cite{kato}
\begin{displaymath}
\det(I+AR),
\end{displaymath}
where $A$ and $R$ are $n\times n$ matrices. If $R$ is non-singular and $A$ is of low rank~$m$, this determinant is effectively of dimension $m\times m$, which can be seen by changing to a suitable basis. However, the factorised representation~$AR$ is not important here, so we will instead consider $\det(I+\mathcal{M})$, where $\mathcal{M}$ is of low rank. (The general case then follows by observing that $\mathcal{M}=AR$ has the same rank as $A$.)

\begin{proposition}[Weinstein--Aronszajn]
\propositionlabel{WA}
Let $\mathcal{M}$ be an $n\times n$ matrix of rank $m\le n$. Given an orthonormal set $\{f_1,\dots,f_m\}$ of vectors in $\R^n$ that spans the range of $\mathcal{M}$, the Weinstein--Aronszajn determinant can be computed as an $m\times m$ determinant:
\begin{align*}
\det(I+\mathcal{M})&=\det(\delta_{ij}+f_j^\trans \mathcal{M}f_k)_{i,j=1}^m.\\
\intertext{In particular, for $m=1$,}
\det(I+\mathcal{M})&=1+f_1^\trans \mathcal{M}f_1,\\
\intertext{and for $m=2$,}
\det(I+\mathcal{M})&=
\begin{vmatrix}
1+f_1^\trans \mathcal{M}f_1 & f_1^\trans \mathcal{M}f_2 \\
f_2^\trans \mathcal{M}f_1 & 1+f_2^\trans \mathcal{M}f_2 
\end{vmatrix}.
\end{align*}
\end{proposition}

\begin{proof} 
Extend the given set $\{f_1,\dots,f_m\}$ to an orthonormal basis $\{f_1,\dots,f_m,\allowbreak f_{m+1},\dots,f_n\}$ for $\R^n$. We represent $\mathcal{M}$ by a sum of $m$ outer products as $\mathcal{M}=\sum_{a=1}^mf_a\,\tilde{f}_a^\trans$ with $\tilde{f}_a=\mathcal{M}^\trans f_a$. In the standard basis $\{e_1,\dots,e_n\}$, the corresponding matrix is then $F^\trans \mathcal{M}F=\sum_{a=1}^me_a\,\tilde{e}_a^\trans$, where $F=(f_1,\dots,f_n)$ and $\tilde{e}_a=F^\trans\tilde{f}_a$. Thus in $F^\trans \mathcal{M}F$ all elements in rows $m+1$, \dots, $n$ vanish. This implies the block form
\begin{displaymath}
\det(I+ \mathcal{M})=
\det(I+F^\trans \mathcal{M}F)=
\begin{vmatrix}
\underline{\delta_{ij}+f_i^\trans \mathcal{M}f_j} \\
\hfill \makebox[0pt]{$0$} \hfill \big\vert 
\hfill \makebox[0pt]{$\displaystyle\delta_{k\ell}$} \hfill
\end{vmatrix},
\end{displaymath}
which shows that expansion of this determinant according to the last row results in a new, lower-dimensional determinant of the same form. This process can thus be repeated to eventually find an $m\times m$ determinant.
\end{proof}

We now apply this proposition to the case when at least one of $G$ and $\Gt$ is quadratic in $q$, that is, at least one of $\alpha$ or $\tilde{\alpha}$ is non-zero.

\begin{theorem}[Cofactor-elliptic coordinates]
\theoremlabel{CPSell}
Let $G(q)=\alpha qq^\trans+q\beta^\trans+\beta q^\trans+\gamma$ and $\Gt(q)=\tilde{\alpha}qq^\trans+q\tilde{\beta}^\trans+ \tilde{\beta}q^\trans+\tilde{\gamma}$, with not both $\alpha$ and $\tilde{\alpha}$ zero, be two elliptic coordinates matrices. Define a vector $B=B(\mu)$ in $\R^n$ and a symmetric $n\times n$ matrix $\Gamma=\Gamma(\mu)$ as
\begin{displaymath}
B(\mu)=\frac{\beta-\mu\tilde{\beta}}{\alpha-\mu\tilde{\alpha}},\quad
\Gamma(\mu)=\frac{\gamma-\mu\tilde{\gamma}}{\alpha-\mu\tilde{\alpha}}.
\end{displaymath}
Choose an orthogonal matrix $S=S(\mu)$ so that $S^\trans(\Gamma-BB^\trans)S$ is diagonal with eigenvalues $\lambda_1=\lambda_1(\mu)$, \dots, $\lambda_{n-1}=\lambda_{n-1}(\mu)$. Define a $\mu$-dependent Euclidean change of reference frame by $Q(q)=S^\trans(q+B)$. 

If all $\lambda_i(\mu)$ are non-zero, the roots $u_1(q)$, \dots, $u_n(q)$ of the polynomial $\det(G-\mu\Gt)$ satisfy the rational equation
\begin{equation}
\label{cpsell}
1+\sum_{i=1}^n\frac{Q_i(q;\mu)^2}{\lambda_i(\mu)}=
\frac{\dG(q)}{(\alpha-\mu\tilde{\alpha})^n}\cdot
\prod_{j=1}^n(u_j-\mu)\bigg/
\prod_{k=1}^n\lambda_k(\mu).
\end{equation}
\end{theorem}

\begin{proof}
In order to use \propositionref{WA}, we rewrite the pencil $G-\mu\Gt$ as 
\begin{multline*}
G-\mu\Gt=(\alpha-\mu\tilde{\alpha})qq^\trans+
(\beta-\mu\tilde{\beta})q^\trans+
q(\beta-\mu\tilde{\beta})^\trans+
\gamma-\mu\tilde{\gamma}=\\
(\alpha-\mu\tilde{\alpha})\bigl[
\bigl(q+B(\mu)\bigr)\bigl(q+B(\mu)\bigr)^\trans+
\Gamma(\mu)-B(\mu)\,B(\mu)^\trans\bigr]=\\
(\alpha-\mu\tilde{\alpha})\bigl[A(q;\mu)+T(\mu)\bigr]=
(\alpha-\mu\tilde{\alpha})\bigl[I+A(q;\mu)\,T(\mu)^{-1}\bigr]T(\mu),
\end{multline*}
where $A=(q+B)(q+B)^\trans$ and $T=\Gamma-BB^\trans$. Using that $G-\mu\Gt=\Gt(X-\mu I)$, we find
\begin{equation}
\label{waell}
\det(I+AT^{-1})=
\frac{\dG}{(\alpha-\mu\tilde{\alpha})^n}\frac{\det(X-\mu I)}{\det T}.
\end{equation}
The left hand side can easily be calculated by invoking \propositionref{WA} for the rank $1$ matrix $AT^{-1}$, and by diagonalizing the symmetric matrix $T$. Indeed, by setting $f_1=(q+B)/\abs{q+B}$, and by choosing an orthogonal matrix $S$ such that $S^\trans TS=\Lambda=\diag(\lambda_1,\dots,\lambda_n)$, we have
\begin{multline*}
\det(I+AT^{-1})=
1+f_1^\trans AT^{-1}f_1=\\
1+(q+B)^\trans T^{-1}(q+B)=
1+\bigl(S^\trans(q+B)\bigr)^\trans\Lambda^{-1}\bigl(S^\trans(q+B)\bigr).
\end{multline*}
Equation \eqref{cpsell} now follows from \eqref{waell}.
\end{proof}

As is indicated in the following remark, the coordinates that satisfy \eqref{cpsell} generalize the classical elliptic coordinates. Since these new coordinates naturally arise in the theory of cofactor pair systems, we call them ``cofactor-elliptic'' coordinates.

\begin{remark}[Elliptic coordinates]
\remarklabel{ell}
If \theoremref{CPSell} is applied with
\begin{displaymath}
G(q)=-qq^\trans+\diag(\lambda_1,\dots,\lambda_n)
\quad\text{and}\quad
\Gt=I,
\end{displaymath}
then $B=0$ and $\Gamma=\diag(\mu-\lambda_1,\dots,\mu-\lambda_n)$, and we recover \propositionref{genell}.
\end{remark}

\begin{example}[Cofactor-elliptic coordinates in $\pmb{\R^2}$]
\examplelabel{HL}
This example has previously been studied by H.~Lundmark, who calculated the relevant expression directly from the pencil $G-\mu\Gt$ by completing squares.

Suppose that
\begin{displaymath}
G(q)=
\begin{pmatrix}
-q_1^2+\lambda_1 & -q_1q_2 \\
-q_1q_2 & -q_2^2+\lambda_2
\end{pmatrix}
\quad\text{and}\quad
\Gt(q)=
\begin{pmatrix}
2\epsilon q_1+(1-\epsilon) & \epsilon q_2 \\
\epsilon q_2 & 1
\end{pmatrix},
\end{displaymath}
where $\lambda_1$, $\lambda_2$ and $\epsilon$ are parameters. The case of one quadratic and one linear elliptic coordinates matrix is in fact the most general, since if both $G$ and $\Gt$ were quadratic, one could consider the pair $G$ and $\Gt-(\tilde{\alpha}/\alpha)G$ instead. When $\epsilon=0$, we have the situation in \remarkref{ell}.

Following the notation of \theoremref{CPSell}, we have $B=(\epsilon\mu,0)^\trans$ and $\Gamma=\diag\bigl((1-\epsilon)\mu-\lambda_1,\mu-\lambda_2\bigr)$, so that
\begin{displaymath}
\Gamma-BB^\trans=
\diag\bigl((1-\epsilon)\mu-\lambda_1-\epsilon^2\mu^2,\mu-\lambda_2\bigr)
\quad\text{and}\quad S=I.
\end{displaymath}
From the theorem we get the rational equation 
\begin{multline*}
1+\frac{(q_1+\epsilon\mu)^2}{(1-\epsilon)\mu-\lambda_1-\epsilon^2\mu^2}
+
\frac{q_2^2}{\mu-\lambda_2}
=\\
\bigl(2\epsilon q_1-\epsilon^2q_2^2+(1-\epsilon)\bigr)
\frac{(u_1-\mu)(u_2-\mu)}
{\bigl((1-\epsilon)\mu-\lambda_1-\epsilon^2\mu^2\bigr)(\mu-\lambda_2)},
\end{multline*}
describing the relation between the separation coordinates $(u_1,u_2)$ and the Cartesian coordinates $(q_1,q_2)$. The coordinate curves of constant $\mu=u$ are quadrics given by \begin{equation}
\label{majhlex}
\frac{(q_1+\epsilon u)^2}{\lambda_1-(1-\epsilon)u+\epsilon^2u^2}
+\frac{q_2^2}{\lambda_2-u}
=1.
\end{equation}
This relation is more complicated to comprehend than the corresponding one for elliptic coordinates. It is the signs of the denominators that determine the kind of quadrics involved. Here we consider two cases: $\epsilon=0$ and $\epsilon=1$.

When $\epsilon=0$, \eqref{majhlex} specializes to
\begin{equation}
\label{ellipses}
\frac{q_1^2}{\lambda_1-u}
+\frac{q_2^2}{\lambda_2-u}
=1.
\end{equation}
We assume that $\lambda_1<\lambda_2$ (there is no qualitative difference in assuming $\lambda_1>\lambda_2$) and study the signs of the denominators as $u$ varies. For $u<\lambda_1$, we have always $(+,+)$ defining a family of confocal ellipses; for $\lambda_1<u<\lambda_2$, we have $(-,+)$ defining a family of confocal hyperbolas; finally for $u>\lambda_2$, we have $(-,-)$ for which there are no real solutions $(q_1,q_2)$ to \eqref{ellipses}. These two families of curves define 1-1 mappings of the region $\{(u_1,u_2)\in\R^2;u_1<\lambda_1<u_2<\lambda_2\}$ onto each quadrant of the $q_1q_2$-plane; therefore we have a coordinate system in the whole plane except on the $q_1$- and $q_2$-axes.

When $\epsilon=1$, \eqref{majhlex} specializes to
\begin{equation}
\label{hlex}
\frac{(q_1+u)^2}{\lambda_1+u^2}
+\frac{q_2^2}{\lambda_2-u}
=1,
\end{equation}
and the picture is more complicated. First of all, if $\lambda_1>0$, then $\lambda_1+u^2>0$, and we have $(+,+)$ if $u<\lambda_2$ and $(+,-)$ if $u>\lambda_2$. Also in this case the equation define a family of ellipses and a family of hyperbolas, but they are not confocal since the curves are being translated along the $q_1$-axis. The curves do not fill the entire $q_1q_2$-plane either, and we can only define a coordinate system in certain regions of the plane. Figure~\ref{fig:epic+5+0} illustrates this situation.

\begin{figure}
\begin{center}
\includegraphics[scale=\scale]{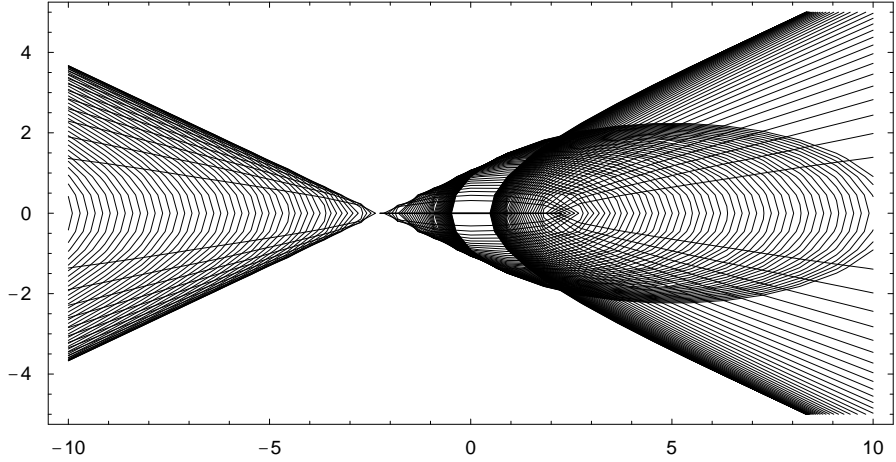}
\caption{Solutions to \eqref{hlex} with $\lambda_1=5$, $\lambda_2=0$ and $-5\le u\le5$.}
\label{fig:epic+5+0}
\end{center}
\end{figure}

The picture is similar when $\lambda_1<0$, except that three or four families of curves are defined as is indicated by the combination of signs in Table~\ref{tab:hlex}. Three different cases occur depending on the relation between $\lambda_2$ and $\pm\sqrt{-\lambda_1}$. They are illustrated in Figures~\ref{fig:epic-5-3}, \ref{fig:epic-5+0} and~\ref{fig:epic-5+3}.

\begin{table}
\begin{center}
\def\dist{\hspace*{3.2em}}
\def\arrow#1#2%
{\xrightarrow[\hspace{2.9em}]{\rule{0mm}{4mm}(#1,#2)}}
\begin{tabular}{|l|c@{}c@{}c@{}c@{}c@{}c@{}c|}
\hline\hline\multicolumn{1}{|c|}{\emph{Case}} &
\multicolumn{7}{c|}{\emph{Signs and curve types along the $u$ axis}} \\ \hline
\hline$\lambda_2\in(-\infty,-\lambda_1^*)$  &
$\arrow++$&$\lambda_2$&$\arrow+-$&
$-\lambda_1^*$&$\arrow--$&$\lambda_1^*$&$\arrow+-$\\
& Ell &\dist& Hyp &\dist& Void &\dist& Hyp \\
\hline$\lambda_2\in(-\lambda_1^*,\lambda_1^*)$ &
$\arrow++$&$-\lambda_1^*$&$\arrow-+$&
$\lambda_2$&$\arrow--$&$\lambda_1^*$&$\arrow+-$\\
& Ell && Hyp && Void && Hyp \\ 
\hline$\lambda_2\in(\lambda_1^*,+\infty)$ &
$\arrow++$&$-\lambda_1^*$&$\arrow-+$&
$\lambda_1^*$&$\arrow++$&$\lambda_2$&$\arrow+-$\\
& Ell && Hyp && Ell && Hyp \\ 
\hline\hline
\end{tabular}
\caption{Different solution types to \eqref{hlex} with $\lambda_1<0$. Notation $\lambda_1^*=\sqrt{-\lambda_1}$.}
\label{tab:hlex}
\end{center}
\end{table}

\begin{figure}
\begin{center}
\includegraphics[scale=\scale]{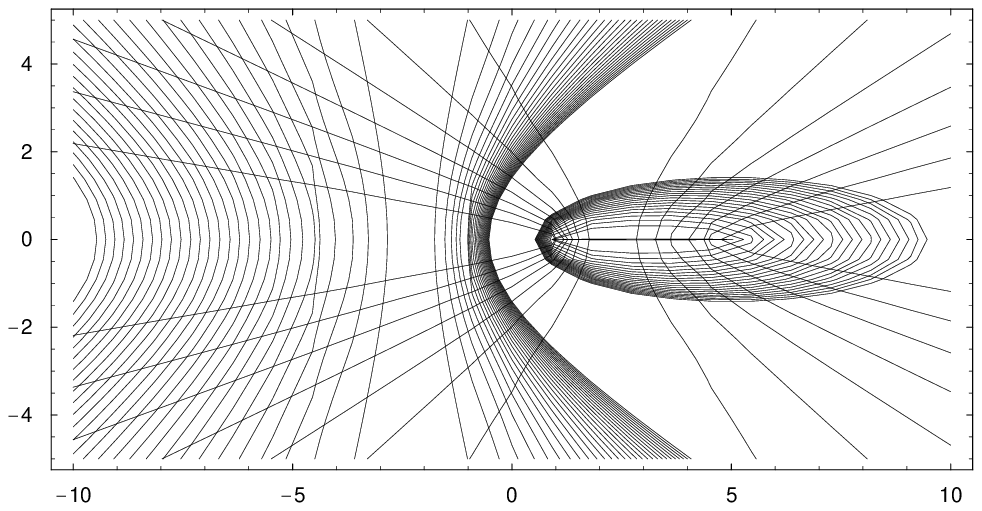}
\caption{Solutions to \eqref{hlex} with $\lambda_1=-5$, $\lambda_2=-3$ and $-5\le u\le5$.}
\label{fig:epic-5-3}
\end{center}
\end{figure}

\begin{figure}
\begin{center}
\includegraphics[scale=\scale]{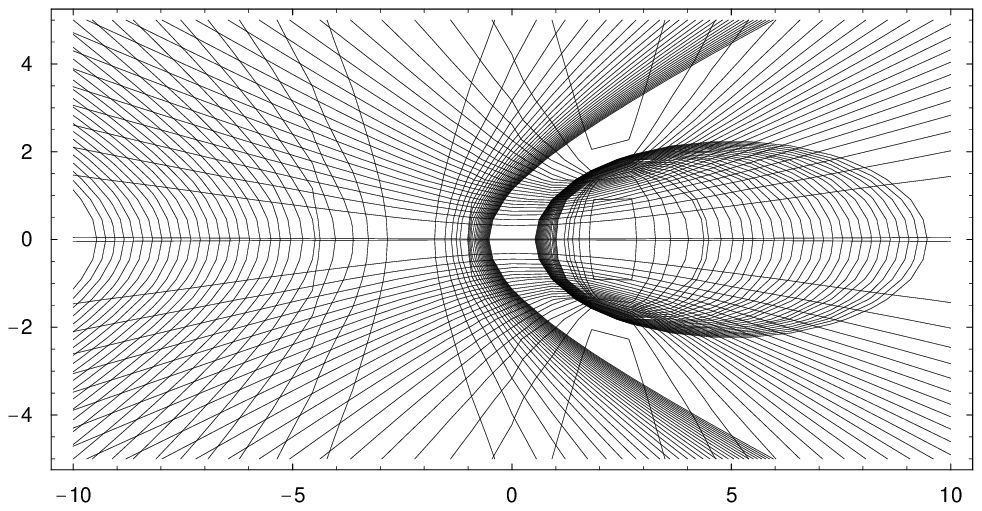}
\caption{Solutions to \eqref{hlex} with $\lambda_1=-5$, $\lambda_2=0$ and $-5\le u\le5$.}
\label{fig:epic-5+0}
\end{center}
\end{figure}

\begin{figure}
\begin{center}
\includegraphics[scale=\scale]{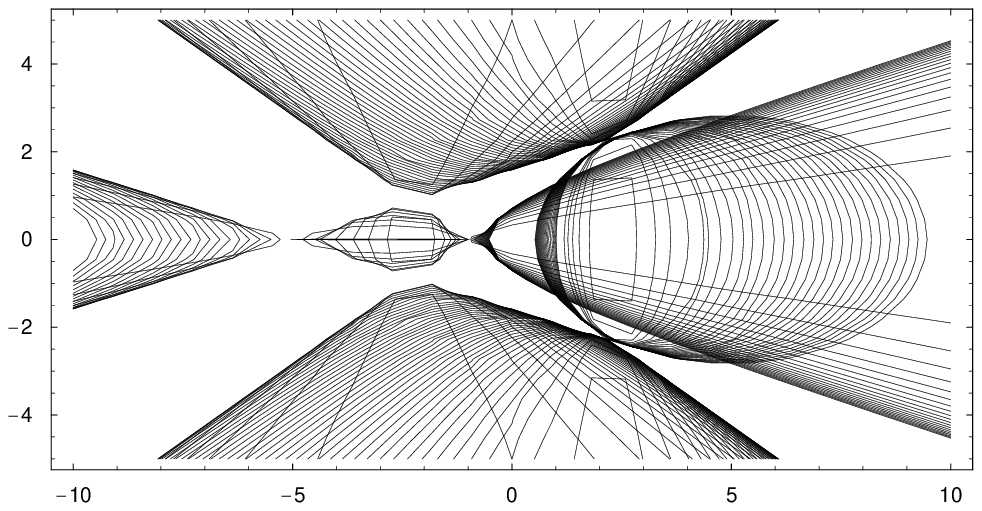}
\caption{Solutions to \eqref{hlex} with $\lambda_1=-5$, $\lambda_2=3$ and $-5\le u\le5$.}
\label{fig:epic-5+3}
\end{center}
\end{figure}
\end{example}

\begin{example}[Cofactor-elliptic coordinates in $\pmb{\R^3}$]
We now extend \exampleref{HL} in the case $\epsilon=1$ to three dimensions. We take
\begin{displaymath}
G(q)=
\begin{pmatrix}
-q_1^2+\lambda_1 & -q_1q_2 & -q_1q_3 \\
-q_1q_2 & -q_2^2+\lambda_2 & -q_2q_3 \\
-q_1q_3 & -q_2q_2 & -q_3^2+\lambda_3 
\end{pmatrix}
\quad\text{and}\quad
\Gt(q)=
\begin{pmatrix}
2q_1 & q_2 & q_3 \\
q_2 & 1 & 0Ê\\
q_3 & 0 & 0
\end{pmatrix},
\end{displaymath}
where $\lambda_1$, $\lambda_2$, $\lambda_3$ are parameters. As before, we find $B=(\mu,0,0)^\trans$ and $\Gamma=\diag(-\lambda_1,-\lambda_2+\mu,-\lambda_3)$ so that
\begin{displaymath}
\Gamma-BB^\trans=
\diag(-\lambda_1-\mu^2,-\lambda_2+\mu,-\lambda_3)
\quad\text{and}\quad S=I.
\end{displaymath}
We get the rational equation 
\begin{displaymath}
1+\frac{(q_1+\mu)^2}{-\lambda_1-\mu^2}
+\frac{q_2^2}{-\lambda_2+\mu}
+\frac{q_3^2}{-\lambda_3}
=q_3^2
\frac{(u_1-\mu)(u_2-\mu)(u_3-\mu)}
{(-\lambda_1-\mu^2)(-\lambda_2+\mu)(-\lambda_3)},
\end{displaymath}
describing the relation between the separation coordinates $(u_1,u_2,u_3)$ and the Cartesian coordinates $(q_1,q_2,q_3)$. The curves of constant $\mu=u$ satisfy
\begin{displaymath}
\frac{(q_1+u)^2}{\lambda_1+u^2}
+\frac{q_2^2}{\lambda_2-u}
+\frac{q_3^2}{\lambda_3}
=1.
\end{displaymath}
If $\lambda_3>0$, there are always real solutions $(q_1,q_2,q_3)$ to this equation for any given $u$. The character of the curves are determined by the relation between $\lambda_1$, $\lambda_2$ and $u$, as explained in \exampleref{HL}.
\end{example}

\subsection{Cofactor-parabolic coordinates}

We shall now investigate the separation coordinates in the case when none of $G$ and $\Gt$ are quadratic, but at least one is linear, in $q$. That is, we assume here that $\alpha=\tilde{\alpha}=0$ and that one of $\beta$ and $\tilde{\beta}$ is non-zero. As in the case of \theoremref{CPSell}, we start with \propositionref{WA}. In order to formulate and prove a similar theorem, we need the following fact about ``partial diagonalisation'' of symmetric matrices.

\begin{proposition}[Partial diagonalisation]
\propositionlabel{almostdiagonal}
Let $\Gamma$ be a symmetric $n\times n$ matrix, and let $\{B_{m+1},\dots,B_n\}$ be an orthonormal set of vectors in $\R^n$. There exists an orthogonal matrix~$S$ with columns $S_j=B_j$ for $j=m+1,\dots,n$, such that $S^\trans\Gamma S$ attains the block form
\begin{displaymath}
\begin{pmatrix}
D & C_1 \\ C_1^\trans & C_2
\end{pmatrix},
\end{displaymath}
where $D$ is a diagonal $m\times m$ matrix, $C_1$ is an $m\times(n-m)$ matrix and $C_2$ is a symmetric $(n-m)\times(n-m)$ matrix. In particular, if $m=n-1$, there is a vector $c$ such that
\begin{displaymath}
S^\trans\Gamma S=
\diag(\lambda_1,\dots,\lambda_{n-1},0)+
e_nc^\trans+ce_n^\trans,
\quad e_n=(0,\dots,0,1)^\trans.
\end{displaymath}
\end{proposition}

\begin{proof}
Take any orthogonal matrix $\tilde{S}$ with columns $\tilde{S}_j=B_j$ for $j=m+1,\dots,n$. Then
\begin{displaymath}
\tilde{S}^\trans\Gamma\tilde{S}= 
\begin{pmatrix}
\tilde{D} & \tilde{C}_1 \\ \tilde{C}_1^\trans & \tilde{C}_2
\end{pmatrix}
\end{displaymath}
for some matrices $\tilde{D}$, $\tilde{C}_1$ and $\tilde{C}_2$. Choose an orthogonal $m\times m$ matrix $P$ such that $P^\trans\tilde{D}P=D$ and set
\begin{displaymath}
S=\tilde{S}
\begin{pmatrix}
P & 0 \\ 0 & I
\end{pmatrix}.
\end{displaymath}
Then $S$ has the stated last columns, and
\begin{displaymath}
S^\trans\Gamma S=
\begin{pmatrix}
P^\trans & 0 \\ 0 & I
\end{pmatrix}
\begin{pmatrix}
\tilde{D} & \tilde{C}_1 \\ \tilde{C}_1^\trans & \tilde{C}_2
\end{pmatrix}
\begin{pmatrix}
P & 0 \\ 0 & I
\end{pmatrix}=
\begin{pmatrix}
D & C_1 \\ C_1^\trans & C_2
\end{pmatrix}
\end{displaymath}
where $C_1=P^\trans\tilde{C}_1$ and $C_2=\tilde{C}_2$.
\end{proof}

\begin{theorem}[Cofactor-parabolic coordinates]
\theoremlabel{CPSpar}
Let $G(q)=q\beta^\trans+\beta q^\trans+\gamma$ and $\Gt(q)=q\tilde{\beta}^\trans+ \tilde{\beta}q^\trans+\tilde{\gamma}$, with not both $\beta$ and $\tilde{\beta}$ zero, be two elliptic coordinates matrices. Define a vector $B=B(\mu)$ in $\R^n$ and a symmetric $n\times n$ matrix $\Gamma=\Gamma(\mu)$ as
\begin{displaymath}
B(\mu)=\frac{\beta-\mu\tilde{\beta}}{\abs{\beta-\mu\tilde{\beta}}},\quad
\Gamma(\mu)=\frac{\gamma-\mu\tilde{\gamma}}{\abs{\beta-\mu\tilde{\beta}}}.
\end{displaymath}
Take $S=S(\mu)$, $\lambda_1=\lambda_1(\mu)$, \dots, $\lambda_{n-1}=\lambda_{n-1}(\mu)$ and $c=c(\mu)$ from \propositionref{almostdiagonal} applied with $\Gamma$ and $\{B\}$. Define a $\mu$-dependent Euclidean change of reference frame by $Q(q)=S^\trans q+c$. 

If all $\lambda_i(\mu)$ are non-zero, the roots $u_1(q)$, \dots, $u_n(q)$ of the polynomial $\det(G-\mu\Gt)$ satisfy the rational equation
\begin{equation}
\label{cpspar}
2Q_n(q;\mu)-\sum_{i=1}^{n-1}\frac{Q_i(q;\mu)^2}{\lambda_i(\mu)}=
\frac{\dG(q)}{\abs{\beta-\mu\tilde{\beta}}^n}\cdot
\prod_{j=1}^n(u_j-\mu)\bigg/
\prod_{k=1}^{n-1}\lambda_k(\mu).
\end{equation}
\end{theorem}

\begin{proof}
The proof is similar to that of \theoremref{CPSell}, but more technical. We rewrite the pencil $G-\mu\Gt$ as 
\begin{multline*}
G-\mu\Gt=(\beta-\mu\tilde{\beta})q^\trans+
q(\beta-\mu\tilde{\beta})^\trans+
\gamma-\mu\tilde{\gamma}=\\
\abs{\beta-\mu\tilde{\beta}}\bigl(
B(\mu)(q+V)^\trans+(q+V)B(\mu)^\trans+
\Gamma(\mu)-B(\mu)\,V^\trans-V\,B(\mu)^\trans\bigr)=\\
\abs{\beta-\mu\tilde{\beta}}\bigl(A(q;\mu)+T(\mu)\bigr)=
\abs{\beta-\mu\tilde{\beta}}\bigl(I+A(q;\mu)\,T(\mu)^{-1}\bigr)T(\mu),
\end{multline*}
where $A=B(q+V)^\trans+(q+V)B^\trans=Bq_V^\trans+q_VB^\trans$, which defines $q_V$, and where $T=\Gamma-BV^\trans-VB^\trans$ and $V$ is an as yet undetermined vector. We get
\begin{equation}
\label{wapar}
\det(I+AT^{-1})=
\frac{\dG}{\abs{\beta-\mu\tilde{\beta}}^n}\frac{\det(X-\mu I)}{\det T}.
\end{equation}
The left hand side can be calculated by invoking \propositionref{WA} for the rank $2$ matrix $AT^{-1}$ by setting $f_1=B$ and $f_2=\bigl(q_V-(q_V^\trans B)B\bigr)/a$ where $a^2=q_V^\trans q_V-(q_V^\trans B)^2$. A straightforward calculation gives
\begin{displaymath}
\begin{vmatrix}
1+f_1^\trans AT^{-1}f_1 & f_1^\trans AT^{-1}f_2 \\
f_2^\trans AT^{-1}f_1 & 1+f_2^\trans AT^{-1}f_2 
\end{vmatrix}=
(1+q_V^\trans T^{-1}B)^2-
(q_V^\trans T^{-1}q_V)(B^\trans T^{-1}B).
\end{displaymath}

This is the point were it should be apparent that an ordinary diagonalisation of $T$ does not help; the problem is the product $T^{-1}B$, which would attain the form $S\Lambda^{-1}S^\trans B$, which is not very useful unless we can simplify the expression $S^\trans B$. However, \propositionref{almostdiagonal} gives us an orthogonal matrix~$S$ having $B$ as last column, and therefore $S^\trans B=e_n$, which proves helpful. 

The matrix $S$ provides the decomposition $S^\trans\Gamma S=\diag(\lambda_1,\dots,\lambda_{n-1},0)+e_nc^\trans+ce_n^\trans$, which implies
\begin{displaymath}
S^\trans TS=\diag(\lambda_1,\dots,\lambda_{n-1},0)+
e_n(c-S^\trans V)^\trans+(c-S^\trans V)e_n^\trans.
\end{displaymath}
We now want this to be diagonal with non-zero diagonal elements, since we already have assumed that $T^{-1}$ exists. Therefore we choose $V$ so that $c-S^\trans V=\onehalf e_n$, that is, we choose $V=Sc-\onehalf B$. Thus
\begin{displaymath}
S^\trans TS=\Lambda=\diag(\lambda_1,\dots,\lambda_{n-1},1),
\end{displaymath}
and the above expression for $\det(I+AT^{-1})$ becomes, with $\tilde{Q}=S^\trans q_V$,
\begin{displaymath}
(1+\tilde{Q}^\trans\Lambda^{-1}e_n)^2-
(\tilde{Q}^\trans\Lambda^{-1}\tilde{Q})(e_n^\trans\Lambda^{-1}e_n)=
1+2\tilde{Q}_n-\sum_{i=1}^{n-1}\frac{\tilde{Q}_i^2}{\lambda_i}.
\end{displaymath}
Changing to the notation in the formulation of the theorem, $\tilde{Q}_i=Q_i$ for $i=1,\dots,n-1$ and $1+2\tilde{Q}_n=2Q_n$, which shows that \eqref{wapar} gives~\eqref{cpspar}.
\end{proof}

As in the elliptic case, the coordinates defined by \eqref{cpspar} will be called ``cofactor-parabolic.''

\begin{remark}[Parabolic coordinates]
If \theoremref{CPSpar} is applied with
\begin{displaymath}
G(q)=e_nq^\trans+qe_n^\trans+\diag(\lambda_1,\dots,\lambda_{n-1},0)
\quad\text{and}\quad
\Gt=I,
\end{displaymath}
then $B=e_n$ and $\Gamma=\diag(\lambda_1-\mu,\dots,\lambda_{n-1}-\mu,-\mu)$, so that $S=I$ and $c=-\onehalf\mu e_n$. This gives $Q=q-\onehalf\mu e_n$, that is, $Q_i=q_i$ for $i=1,\dots,n-1$ and $2Q_n=2q_n-\mu$. \propositionref{genpar} is thus recovered.
\end{remark}

\begin{example}[Cofactor-parabolic coordinates in $\pmb{\R^2}$]
Suppose that
\begin{displaymath}
G(q)=
\begin{pmatrix}
\lambda & q_1 \\
q_1 & 2q_2
\end{pmatrix}
\quad\text{and}\quad
\Gt(q)=
\begin{pmatrix}
2q_1 &  q_2 \\
q_2 & 0
\end{pmatrix},
\end{displaymath}
where $\lambda$ is a parameter. Following the notation of \theoremref{CPSpar}, we have $B=(-\mu,1)^\trans/N(\mu)$ and $\Gamma=\diag(\lambda,0)/N(\mu)$, where $N(\mu)=\sqrt{1+\mu^2}$. In \propositionref{almostdiagonal} we can take
\begin{displaymath}
S=
\frac{1}{N(\mu)}
\begin{pmatrix}
1 & -\mu \\
\mu & 1
\end{pmatrix}
\end{displaymath}
so that
\begin{displaymath}
S^\trans\Gamma S=
\frac{1}{N(\mu)^3}
\begin{pmatrix}
\lambda & -\mu\lambda \\
-\mu\lambda & \mu^2\lambda 
\end{pmatrix},
\end{displaymath}
from which $\lambda_1=\lambda/N(\mu)^3$ and $c=(-\mu\lambda,\onehalf\mu^2\lambda)^\trans/N(\mu)^3$. The theorem gives the rational equation
\begin{multline*}
2\frac{q_2-\mu q_1}{N(\mu)}
+\frac{\mu^2\lambda}{N(\mu)^3}
-\frac{N(\mu)^3}{\lambda}
\biggl(\frac{q_1+\mu q_2}{N(\mu)}-\frac{\mu\lambda}{N(\mu)^3}\biggr)^2
=\\
\frac{-q_2^2}{N(\mu)^2}\cdot
\frac{N(\mu)^3(u_1-\mu)(u_2-\mu)}{\lambda}.
\end{multline*}
For a fixed $\mu=u$, we thus have
\begin{displaymath}
2\frac{q_2-u q_1}{N(u)}
+\frac{u^2\lambda}{N(u)^3}
-\frac{N(u)^3}{\lambda}
\biggl(\frac{q_1+uq_2}{N(u)}-\frac{u\lambda}{N(u)^3}\biggr)^2=0,
\end{displaymath}
which after multiplication by $N(u)$ and expansion of the square simplifies to
\begin{equation}
\label{parex}
2q_2-\frac{(q_1+uq_2)^2}{\lambda}=0.
\end{equation}

This equation defines a fan of parabolas centered at the origin, filling the entire upper half plane when $\lambda>0$, and similarly in the lower half plane when $\lambda<0$. This is illustrated in Figure~\ref{fig:ppic+1}. 

Equation~\eqref{parex} allows for a global definition of the coordinates, as is easily seen by solving the equation for $u$. Indeed, for a fixed $\lambda$ we can take
\begin{displaymath}
u_1(q)=\frac{-q_1+\sqrt{2\lambda q_2}}{q_2}
\quad\text{and}\quad
u_2(q)=\frac{-q_1-\sqrt{2\lambda q_2}}{q_2}
\end{displaymath}
everywhere in the region $\{(q_1,q_2)\in\R^2;\lambda q_2>0\}$.

\begin{figure}
\begin{center}
\includegraphics[scale=\scale]{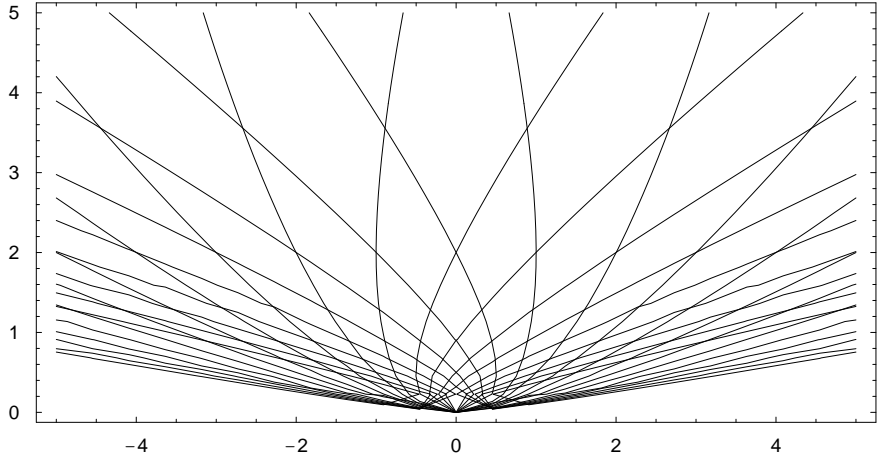}
\caption{Solutions to \eqref{parex} with $\lambda=1$ and $-5\le u\le5$.}
\label{fig:ppic+1}
\end{center}
\end{figure}
\end{example}

\subsection{The functional independence assumption}\sectionlabel{functionalindep}

We complete this study of the separation coordinates $u_k$ by giving a formula that can be used to investigate whether the functions $u_k(q)$ are functionally independent without having to calculate them explicitly. Recall that they are functionally independent if $\det\Jac\ne0$, where $J=(\grad u_1, \dots, \grad u_n)$ is the previously defined Jacobian. The formula is formulated in terms of the coefficients in the polynomial $\det(X-\mu I)$, but if only $G$ and $\Gt$ are given it is not necessary to calculate $X=\Gt^{-1}G$, since $\det(X-\mu I)=\det(G-\mu\Gt)/\det\Gt$.

\begin{proposition}[A formula for the Jacobian $\pmb{\Jac}$]
Let $a_0$, \dots, $a_n$ be the coefficients in the characteristic polynomial of $X$ defined by
\begin{displaymath}
\det(X-\mu I)=\sum_{k=0}^na_{n-k}\,(-\mu)^k,
\end{displaymath}
and let $u_k$ be the roots of this polynomial. Then
\begin{equation}
\label{jacformula}
\det(\grad a_1, \dots, \grad a_n)=
\det(\grad u_1, \dots, \grad u_n)\cdot
\prod_{j<k}(u_j-u_k).
\end{equation}
\end{proposition}

\begin{proof}
Since $a_k=\sigma_k(u)$ for all $k$, the chain rule and \eqref{derivativeofsymmetric} imply
\begin{displaymath}
\grad a_k=\sum_j\frac{\partial\sigma_k(u)}{\partial u_j}\grad u_j=
\sum_j\sigmacheck{k-1}{j}\,\grad u_j.
\end{displaymath}
These relations can be put into matrix form as
\begin{displaymath}
(\grad a_1, \dots, \grad a_n)=
(\grad u_1, \dots, \grad u_n)\cdot
 \begin{pmatrix}
        {{\sigmacheck01}} &
        \dots &
        {{\sigmacheck{n-1}1}} \\
        \hdotsfor{3} \\
        {{\sigmacheck0n}} & 
    \dots &
        {{\sigmacheck{n-1}n}}
 \end{pmatrix}.
\end{displaymath}
This implies \eqref{jacformula}, for the matrix $\bigl(\sigmacheck{j}k\bigr)$ is the transpose of the matrix \eqref{sigmacheckmatrix}, which according to \appendixref{sigma} has determinant $\prod_{j<k}(u_j-u_k)$.
\end{proof}

\begin{remark}[An a priori functional independence check]
It is possible to use the formula also to check if the quadratic first integrals defined by a cofactor pair system are functionally independent. Indeed, in view of the separability theorem, they are functionally independent if the functions~$u_k(q)$ are, and the formula shows that it suffices to check the coefficients~$a_k(q)$. 
\end{remark}

\appendix

\section{Symmetric polynomials, etc.}
\appendixlabel{sigma}

In this paper we frequently encounter two types of elementary symmetric polynomials. These are~$\sigma_i(u)$, which are defined by
\begin{equation}
\label{symmetricdefinition}
\prod_{i=1}^n (z+u_i) = \sum_{i=0}^n z^{n-i}\,\sigma_i(u),
\end{equation}
and $\sigmacheck{i}j$, which are defined by
\begin{equation}
\label{checksymmetricdefinition}
\prod_{i\ne j} (z+u_i) = 
\sum_{i=0}^{n-1}z^{n-1-i}\,\sigmacheck{i}j.
\end{equation}
The symbol $\check{u}_j$ indicates that $\sigmacheck{i}j$ depends on all $u_1$, \dots, $u_n$ except $u_j$. We also extend these definitions to include $\sigma_{-1}(u)=0$ and $\sigmacheck{-1}j=0$.

There are two sets of identities that provide a connection between these polynomials. The first is
\begin{equation}
\label{derivativeofsymmetric}
\frac{\partial}{\partial u_j}\sigma_i(u)=\sigmacheck{i-1}j,
\quad i=0,\dots,n,
\end{equation}
which follows by comparing the derivative of \eqref{symmetricdefinition} with \eqref{checksymmetricdefinition}. The second is
\begin{equation}
\label{decomposesymmetric}
\sigma_i(u)=u_j\,\sigmacheck{i-1}j+\sigmacheck{i}j,
\quad i=0,\dots,n,
\end{equation}
which follows by expanding the product in the left hand side of \eqref{symmetricdefinition}, except for the factor $(z+u_j)$, using \eqref{checksymmetricdefinition}.

Closely connected with these polynomials are the products
\begin{displaymath}
 U'(u_k)=\left.\frac{d}{dz}\right|_{z=u_k}
 \prod_{i=1}^n(z-u_i)=\prod_{i\ne k}(u_k-u_i).
\end{displaymath}
They arise in the inverse
\begin{equation}
 \label{vandermondeinverse}
 \mathcal{V}^{-1}=
 \begin{pmatrix}
        {{\sigmacheck01}/{U'(u_1)}} &
        \dots &
        {{\sigmacheck0n}/{U'(u_n)}} \\
        \hdotsfor{3} \\
        {{\sigmacheck{n-1}1}/{U'(u_1)}} & 
    \dots &
        {{\sigmacheck{n-1}n}/{U'(u_n)}}
 \end{pmatrix}
\end{equation}
of the Vandermonde matrix
\begin{equation}
 \label{vandermonde}
 \mathcal{V}=
 (-1)^{n+1}
 \begin{pmatrix}
        (-u_1)^{n-1} &
        \dots &
        (-u_1)^0 \\
        \hdotsfor{3} \\
        (-u_n)^{n-1} &
        \dots &
        (-u_n)^0 \\
 \end{pmatrix}.
\end{equation}
By using \eqref{checksymmetricdefinition}, it is easy to check that the two matrices are each others inverses: the element in position $(j,k)$ in the matrix product $\mathcal{V}\cdot\mathcal{V}^{-1}$ is
\begin{multline*}
 \sum_{i=1}^n(-1)^{n+1}(-u_j)^{n-i}\,
 \frac{\sigmacheck{i-1}k}{U'(u_k)}=\\
 \frac{(-1)^{n+1}}{U'(u_k)}
 \sum_{i=0}^{n-1}(-u_j)^{n-1-i}\,\sigmacheck{i}k=
 \frac{(-1)^{2n}}{U'(u_k)}
 \prod_{i\ne k}(u_j-u_i)=\delta_{jk}.
\end{multline*}

We recall the well known fact $\det\mathcal{V}=\prod_{j<k}(u_j-u_k)$. We also need the determinant of the matrix
\begin{equation}
 \label{sigmacheckmatrix}
 \begin{pmatrix}
        {{\sigmacheck01}} &
        \dots &
        {{\sigmacheck0n}} \\
        \hdotsfor{3} \\
        {{\sigmacheck{n-1}1}} & 
    \dots &
        {{\sigmacheck{n-1}n}}
 \end{pmatrix}
 =\mathcal{V}^{-1}\cdot\diag\bigl(U'(u_1),\dots,U'(u_n)\bigr),
\end{equation}
which consequently can be calculated as
\begin{displaymath}
\det\bigl(\sigmacheck{j}k\bigr)=
\prod_{j=1}^nU'(u_j)\bigg/
\prod_{j<k}(u_j-u_k)=
\prod_{j<k}(u_j-u_k).
\end{displaymath}

\section{The Levi-Civita equations}

In the proof of \propositionref{stackelpotential}, we need an explicit expression for the solution to the system $\partial^2(u_i-u_j)V/\partial u_iÊ\partial u_j =0$ of PDEs for $V$. The solution is obtained by considering a special case of the following lemma, which deals with a system of PDEs first derived by T.~Levi-Civita. 

In the formulation of the lemma, we use the concept of a St\"ackel matrix. A St\"ackel matrix is a non-singular matrix $\Phi=\bigl(\phi_{ij}(u_i)\bigr)$, $\det\Phi\ne0$, whose $i$:th row only depends on $u_i$, like, for instance, the Vandermonde matrix~\eqref{vandermonde}.

\begin{lemma}
Suppose that $\Phi$ is a St\"ackel matrix with inverse $\Psi=(\psi_{ij})$ having non-vanishing elements in the first row. The solution of the system of $\tbinom{n}{2}$ PDEs
\begin{equation}
\label{LCV}
\frac{\partial^2 V}{\partial u_i \partial u_j}-
\frac{\partial\log\psi_{1j}}{\partial u_i}
\frac{\partial V}{\partial u_j}-
\frac{\partial\log\psi_{1i}}{\partial u_j}
\frac{\partial V}{\partial u_i}=0\quad(i\ne j)
\end{equation}
is
\begin{equation}
\label{SM}
V(u)=\sum\psi_{1k}\,f_k(u_k)
\end{equation}
where $f_i$ are some functions of one variable.
\end{lemma}

\begin{proof}
We will use the Levi-Civita separability condition~\cite{lc,sb} to find the solution of \eqref{LCV}. This condition states that a Hamiltonian $H(u,s)$ on $\R^{2n}$ with coordinates $(u,s)$ is separable, that is, its corresponding Hamilton--Jacobi equation $H(u,\partial_uS)=E$ admits a complete separated solution $S=\sum S_k(u_k)$, if and only if the equations
\begin{displaymath}
\partial_{ij}H\,\partial^iH\,\partial^jH-
\partial_j^iH\,\partial_iH\,\partial^jH+
\partial^{ij}H\,\partial_iH\,\partial_jH-
\partial_i^jH\,\partial^iH\,\partial_jH=0
\end{displaymath}
are satisfied identically with respect to $(u,s)$ for all distinct $i$, $j=1$, \dots, $n$. Here we write $\partial_i$ for $\partial/\partial u_i$ and $\partial^i$ for $\partial/\partial s_i$. When this condition is applied to the Hamiltonian $H=\sum\psi_{1k}s_k-V$, we find
\begin{multline*}
\sum(\psi_{1i}\,\psi_{1j}\,\partial_{ij}\psi_{1k}
-\psi_{1i}\,\partial_i\psi_{1j}\,\partial_j\psi_{1k}
-\psi_{1j}\,\partial_j\psi_{1i}\,\partial_i\psi_{1k})s_k-\\
(\psi_{1i}\,\psi_{1j}\,\partial_{ij}V
-\psi_{1i}\,\partial_i\psi_{1j}\,\partial_jV
-\psi_{1j}\,\partial_j\psi_{1i}\,\partial_iV)=0.
\end{multline*}
The vanishing of the terms involving $V$ is equivalent to \eqref{LCV}. Similarly, the coefficient of $s_k$ vanishes if and only if $\psi_{1k}$ solves \eqref{LCV}; below we will show that this is in fact so. Thus, if $V$ satisfies \eqref{LCV}, there is a separated solution $S=\sum S_k(u_k)$ to the equation $\sum\psi_{1k}\,\partial_kS-V=0$, which implies \eqref{SM} with $f_k(u_k)=\partial_kS_k(u_k)$.

\emph{Claim: $\psi_{1k}$ solves \eqref{LCV}.} To prove this, we introduce the notation $\phi=\det\Phi$ and $\psi^{ij}_{k\ell}=\psi_{ik}\psi_{j\ell}-\psi_{i\ell}\psi_{jk}$. We note that the product $\phi\psi_{ij}$ is independent of $u_j$ since it equals $(-1)^{i+j}$ times the determinant obtained by deleting row $j$ and column $i$ in $\phi$. Likewise, the product $\phi\psi^{ij}_{k\ell}$ is independent of $u_k$ and $u_\ell$ since it equals $(-1)^{i+j+k+\ell}$ times the determinant obtained by deleting rows $k$, $\ell$ and columns $i$, $j$ in $\phi$~\cite{gant}. In the same spirit, we state the identities
\begin{displaymath} 
\frac{1}{\phi\psi_{1i}\psi_{1j}}=
\frac{1}{\phi\psi^{1k}_{ij}}\biggl(
\frac{\phi\psi_{kj}}{\phi\psi_{1j}}-
\frac{\phi\psi_{ki}}{\phi\psi_{1i}}\biggr)
\quad\text{and}\quad
\frac{\psi_{1k}}{\phi\psi_{1i}\psi_{1j}}=
\frac{1}{\phi\psi^{1k}_{ij}}\biggl(
\frac{\phi\psi^{1k}_{kj}}{\phi\psi_{1j}}-
\frac{\phi\psi^{1k}_{ki}}{\phi\psi_{1i}}\biggr),
\end{displaymath}
which imply that 
\begin{displaymath} 
\partial_{ij}\frac{1}{\phi\psi_{1i}\psi_{1j}}=0
\quad\text{and}\quad
\partial_{ij}\frac{\psi_{1k}}{\phi\psi_{1i}\psi_{1j}}=0,
\end{displaymath}
respectively. On the other hand, by Leibniz' rule,
\begin{multline*}
\partial_{ij}\frac{\psi_{1k}}{\phi\psi_{1i}\psi_{1j}}=
\partial_{ij}\biggl(\frac{1}{\phi\psi_{1i}\psi_{1j}}\biggr)\psi_{1k}+\\
\frac{1}{\phi\psi_{1i}\psi_{1j}}\biggl[
\partial_{ij}\psi_{1k}+
\psi_{1i}\,\partial_j\biggl(\frac{1}{\psi_{1i}}\biggr)
\partial_i\psi_{1k}+
\psi_{1j}\,\partial_i\biggl(\frac{1}{\psi_{1j}}\biggr)
\partial_j\psi_{1k}\biggr].
\end{multline*}
The expression within square brackets thus has to vanish, which proves the claim.
\end{proof}

It is easy to check that \eqref{SM} satisfies \eqref{LCV} with arbitrary functions $f_k(u_k)$. One would therefore like to draw the conclusion that \emph{the general solution of \eqref{LCV} is \eqref{SM} with arbitrary functions $f_k(u_k)$,} but it seems very hard to show this. From the above proof also follows that \eqref{LCV} can be put into the suggestive form
\begin{displaymath}
\frac{\partial^2}{\partial u_i \partial u_j}\biggl(
\frac{V}{\phi\psi_{1i}\psi_{1j}}\biggr)=0,
\end{displaymath}
but this does not seem to help in finding the general solution to the equations by direct methods. 

However, by applying the lemma with $\Phi$ being the Vandermonde matrix \eqref{vandermonde}, we establish the following.

\begin{corollary}
\corollarylabel{annoying-system}
The system $\partial^2(u_i-u_j)V/\partial u_iÊ\partial u_j =0$ has the solution $V=\sum f_k(u_k)/U'(u_k)$ where $f_k$ are some functions of one variable.
\end{corollary}


\bibliographystyle{plain}

\addcontentsline{toc}{section}{References}
\bibliography{cofpair}

\begin{thebibliography}{10}

\bibitem{sb}
S.~Benenti.
\newblock Intrinsic characterization of the variable separation in the
  {Hamilton}--{Jacobi} equation.
\newblock {\em J. Math. Phys.}, 38(12):6578--6602, 1997.

\bibitem{sb-cc-gr}
S.~Benenti, C.~Chanu, and G.~Rastelli.
\newblock Remarks on the connection between the additive separation of the
  {Hamilton}--{Jacobi} equation and the multiplicative separation of the
  {Schr{\"o}dinger} equation.
\newblock {\em J. Math. Phys.}, 43(11):5183--5253, 2002.

\bibitem{gant}
F.~R. Gantmacher.
\newblock {\em The Theory of Matrices}.
\newblock Chelsea, 1977.

\bibitem{jac}
C.~G.~J. Jacobi.
\newblock {\em Vorlesungen {\"u}ber Dynamik}.
\newblock Georg Reimer, Berlin, 1866.
\newblock Jacobi's lectures on dynamics given in {K{\"o}nigsberg} 1842--1843
  published by A. Clebsch.

\bibitem{kato}
T.~Kato.
\newblock {\em Perturbation Theory for Linear Operators}.
\newblock Springer-Verlag, 1984.

\bibitem{lc}
T.~Levi-Civita.
\newblock Sulla integrazione della equazione di {Hamilton}--{Jacobi} per
  separazione di variabili.
\newblock {\em Math. Ann.}, 59:383--397, 1904.

\bibitem{hl}
H.~Lundmark.
\newblock Higher-dimensional integrable {Newton} systems with quadratic
  integrals of motion.
\newblock {\em Stud. Appl. Math.}, 110(3):257--296, 2003.

\bibitem{hl-srw}
H.~Lundmark and S.~Rauch-Wojciechowski.
\newblock Driven {Newton} equations and separable time-dependent potentials.
\newblock {\em J. Math. Phys.}, 43(12):6166--6194, 2002.

\bibitem{fm-gf-mp}
F.~Magri, G.~Falqui, and M.~Pedroni.
\newblock The method of {P}oisson pairs in the theory of nonlinear {PDE}s.
\newblock Technical Report 135/1999/FM, SISSA.
\newblock arXiv:nlin.SI/0002009.

\bibitem{km-mb}
K.~Marciniak and M.~B{\l}aszak.
\newblock Separation of variables in quasi-potential systems of bi-cofactor
  form.
\newblock {\em J. Phys. A: Math. Gen.}, 35:2947--2964, 2002.

\bibitem{srw-km-hl}
S.~Rauch-Wojciechowski, K.~Marciniak, and H.~Lundmark.
\newblock Quasi-{Lagrangian} systems of {Newton} equations.
\newblock {\em J. Math. Phys.}, 40(12):6366--6398, 1999.

\bibitem{stackann}
P.~St{\"a}ckel.
\newblock Ueber die {B}ewegung eines {P}unktes in einer $n$-fachen
  {M}annigfaltigkeit.
\newblock {\em Math. Ann.}, XLII:537--563.

\bibitem{stackhab}
P.~St{\"a}ckel.
\newblock {\em Ueber die Integration der {H}amilton--{J}acobi'schen
  Differen\-tial\-gleichung mittelst Separation der Variabeln}.
\newblock Habilitations\-schrift. Halle, 1891.

\bibitem{cw-srw}
C.~Waksj{\"o} and S.~Rauch-Wojciechowski.
\newblock How to find separation coordinates for the {Hamilton}--{Jacobi}
  equation: a criterion of separability for natural {Hamiltonian} systems.
\newblock {\em Math. Phys. Anal. Geom.}
\newblock To appear.

\end{thebibliography}

\end{document}